\documentclass[aps,prl,reprint,groupedaddress,showpacs,amsmath,amssymb,ocolumn,floatfix,longbibliography]{revtex4-1}
\usepackage{graphicx}
\usepackage{bm}
\usepackage{color}
\usepackage[normalem]{ulem}
\usepackage{tabularx}
\usepackage{lineno}
\usepackage{microtype}

\usepackage[colorlinks=true]{hyperref} 
\hypersetup{
    unicode=false,          % non-Latin characters 
    pdftoolbar=true,        % show Acrobat
    pdfmenubar=true,        % show Acrobat 
    pdffitwindow=false,     % window fit to page when opened
    pdfstartview={FitH},    % fits the width of the page to the window
    pdftitle={XXX},    % title
    pdfauthor={XXX et al.},     % author
    pdfsubject={},   % subject of the document
    pdfcreator={},   % creator of the document
    pdfproducer={}, % producer of the document
    pdfkeywords={} {} {}, % list of keywords
    pdfnewwindow=true,      % links in new window
    colorlinks=true,       % false: boxed links; true: colored links
    linkcolor=magenta, %red,          % color of internal links (change box color with linkbordercolor)
    citecolor=blue,        % color of links to bibliography
    filecolor=magenta,      % color of file links
    urlcolor=blue           % color of external links
} 
\usepackage{cleveref}

\newcommand{\Rpara}{$R_{\parallel} $}
\newcommand{\Rperp}{$R_{\perp}$}

\newcommand{\Bperp}{$B_{\perp}$}

\usepackage{pifont}

\usepackage{bbm}

\renewcommand{\vec}[1]{\boldsymbol{#1}}

\begin{document}

\title{Coexisting Charge Density Wave and Superconducting Order in Quantizing Magnetic Fields}

\author{Ron Q. Nguyen$^{1,2}$}
\thanks{These authors contributed equally to this work.}
\author{Peiyu Qin$^{1}$}
\thanks{These authors contributed equally to this work.}
\author{Hai-Tian Wu$^{2}$}
\thanks{These authors contributed equally to this work.}
\author{Sparsh Mishra$^{1}$}
\author{Tobias Wolf$^{1}$}
\author{Joseph Roll$^{1}$}
\author{Erin Morissette$^{2}$}
\author{Naiyuan J. Zhang$^{2}$}
\author{Sarah Alkidim$^{2}$}
\author{Kenji Watanabe$^{3}$}
\author{Takashi Taniguchi$^{4}$}
\author{Aaron W. Hui$^{2}$}
\author{Dima E. Feldman$^{2,5}$}
\author{Allan MacDonald$^{1}$}
\author{J.I.A. Li$^{1}$}
\email{jia.li@austin.utexas.edu}

\affiliation{$^{1}$Department of Physics, University of Texas at Austin, Austin, TX 78712, USA}
\affiliation{$^{2}$Department of Physics, Brown University, Providence, RI 02912, USA}
\affiliation{$^{3}$Research Center for Functional Materials, National Institute for Materials Science, 1-1 Namiki, Tsukuba 305-0044, Japan}
\affiliation{$^{4}$International Center for Materials Nanoarchitectonics,
National Institute for Materials Science,  1-1 Namiki, Tsukuba 305-0044, Japan}
\affiliation{$^5$Brown Theoretical Physics Center, Brown University, Providence, Rhode Island 02912, USA}

\date{\today}

\maketitle

\textbf{
Charge density wave (CDW) and superconductivity are both common in strongly interacting electron systems. While CDW order is ubiquitous in both quantum Hall systems and unconventional superconductors, superconductivity is generally suppressed by the strong magnetic fields required for Landau quantization.  Here we investigate the intertwined CDW and superconducting phases of rhombohedral hexalayer graphene (R6G) in a large displacement field, which generates tunable flat band edges, and a strong magnetic field, which generates a manifold of
nearly degenerate Landau levels.  
We find a series of integer quantum Hall effects with Hall conductance quantum numbers that deviate from nearby integer filling factors, an observation that can be explained only by CDW order that mixes many Landau levels. We also find a nearby superconducting phase stabilized by perpendicular magnetic fields and persists deep within the quantum Hall regime. This intertwinement provides new insight into superconductivity in R6G at zero magnetic field. 
}

%Charge density wave (CDW) and superconductivity are both common in strongly interacting electron systems. While CDW order is ubiquitous in both quantum Hall systems and unconventional superconductors, superconductivity is generally suppressed by the strong magnetic fields required for Landau quantization.  Here we investigate the intertwined CDW and superconducting phases of rhombohedral hexalayer graphene (R6G) in a large displacement field, which generates tunable flat band edges, and a strong magnetic field, which generates a manifold of nearly degenerate Landau levels.  CDW order is accompanied by pronounced thermal hysteresis as expected for first-order melting transitions.  Surprisingly, we find a series of strong integer quantum Hall effects at magnetic fields above $\sim 2$T with Hall conductance quantum numbers that deviate strongly from nearby integer filling factors, an observation that can be explained only by CDW order that mixes many Landau levels. We also find a nearby superconducting phase that is stabilized by perpendicular magnetic fields and persists deep within the quantum Hall regime. The CDW and superconducting phases develop on comparable temperature scales and emerge from the same manifold of strongly mixed Landau levels. This observations provides new insight into the interplay between superconductivity and CDW order in R6G at zero magnetic field. 

Charge density wave (CDW) order~\cite{Wigner1934,Gruner1988CDW,Peierls1955CDW,Gorkov2012CDW,Varma1983CDW,Borisenko2008CDW,Zhu2015CDW} has played a central role in two major areas of condensed-matter physics: quantum Hall physics and unconventional superconductivity. 
Under Landau quantization, it gives rise to a rich landscape of correlated quantum Hall phases, including Wigner solids ~\cite{Lozovik1975Wigner,Andrei1988Wigner,Williams1991Wigner,Li1991Wigner,Yoon1999Wigner,Chen2003Wigner,Jang2017Wigner,Tsui2024Wigner,Csathy2007CDW,Goldman1990Wigner}, and bubble ~\cite{Eisenstein2002bubble,Fogler2002stripe,Goerbig2003RIQH,Goerbig2004electronsolid,Xia2004bubble,Deng2012reentrant,Gervais_3rdLL,Liu2012reentrant,Halperin2020FQHE,Chen2019RIQH,Friess2018bubblestripe} and stripe phases ~\cite{Fogler1996stripe,Koulakov1996stripe,Lilly1999stripe,Lilly1999stripe2,Du1999stripe,Pan1999stripe,Fradkin1999stripe,Macdonald2000stripe,Sammon2019stripe,Fradkin2010nematic,Friess2018bubblestripe}. In zero or weak magnetic fields, the interplay between CDW order and superconductivity has remained an 
important theme in the study of unconventional superconductors ~\cite{Berg2009stripe,Fradkin2010nematic,Berg2009PDW,Kivelson2003stripe,Vojta2009stripe,ReviewHighTcNRP2021, Csathy2020nematicity, Csathy2016topbreaksym, Csathy2018interaction}.

\begin{figure*}
\includegraphics[width=1\linewidth]{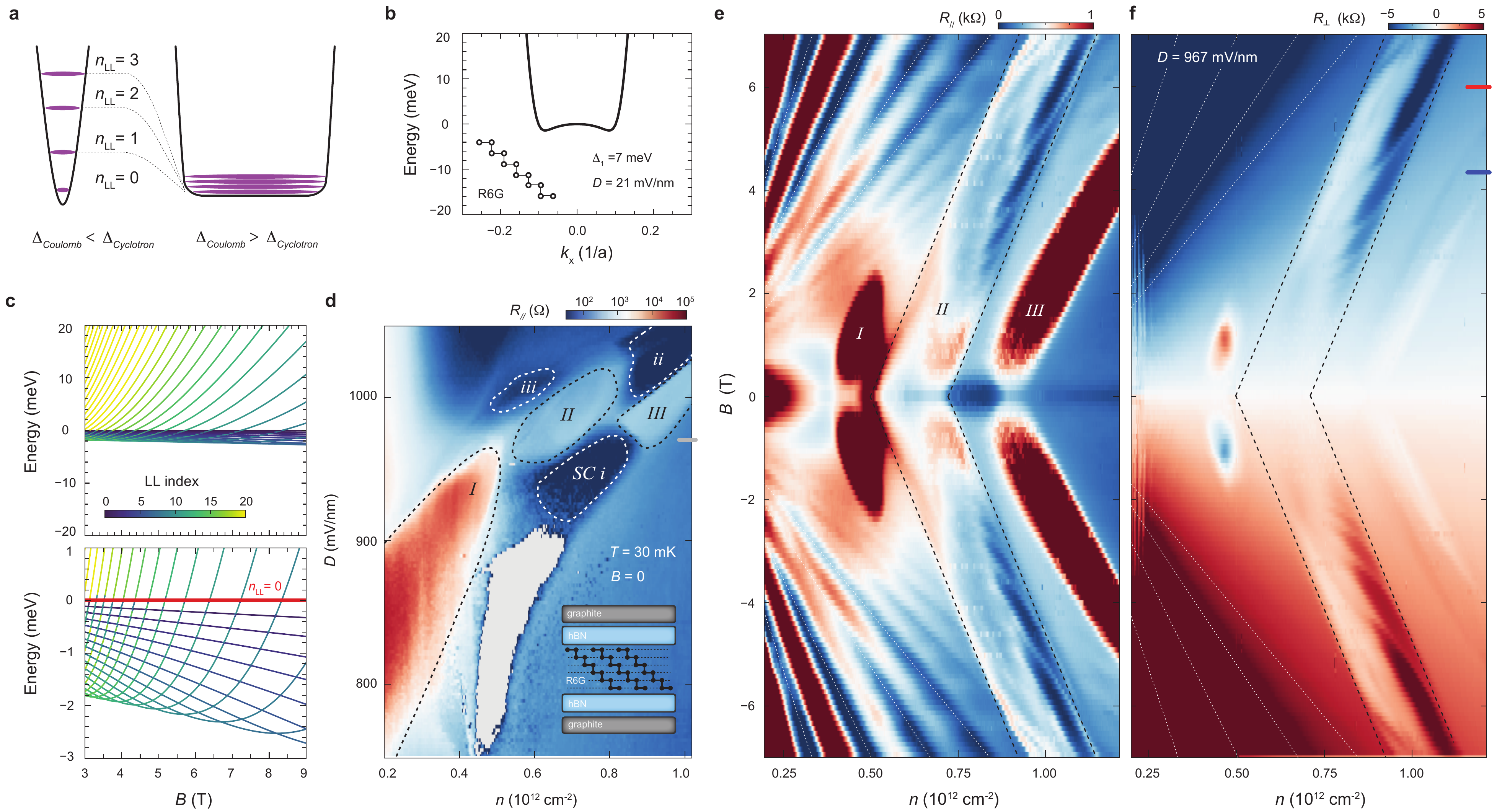}
\caption{
\textbf{LL formation at the flat-band edge.}
(a) Schematic illustration of LL formation in a dispersive band (left) and near a flat-band edge (right). The flat-band edge enhances mixing among nearly degenerate LLs.
(b) Conduction-band dispersion of R6G calculated using Slonczewski--Weiss--McClure (SWMc) parameters $\gamma_0$ and $\gamma_1$ ~\cite{zhangDFT_ABC_trilayer}. This simplified band-structure calculation neglects trigonal warping and other higher-order hopping terms (see Fig.~\ref{band}d for a more realistic calculation).
(c) Landau-level spectrum derived from the band structure shown in (b),  obtained using the continuum model ~\cite{koshinoABC_model}. The bottom panel displays a zoomed-in view of the low-energy edge of the spectrum. 
(d) $n$--$D$ map of the longitudinal resistance \Rpara\ measured at $B_{\perp}=0$. Three superconducting phases (SC\,i--iii) are outlined by white dashed contours, while three interspersed resistive regions (I--III) are outlined by black dashed contours.
(e, f) $n$--\Bperp\ maps of (e) longitudinal resistance, \Rpara, and (f) transverse resistance, \Rperp, measured at a fixed displacement field of $D=967\,\mathrm{mV/nm}$, indicated by the horizontal gray bar in (d).
}
\label{fig1}
\end{figure*}

\begin{figure*}
\includegraphics[width=1\linewidth]{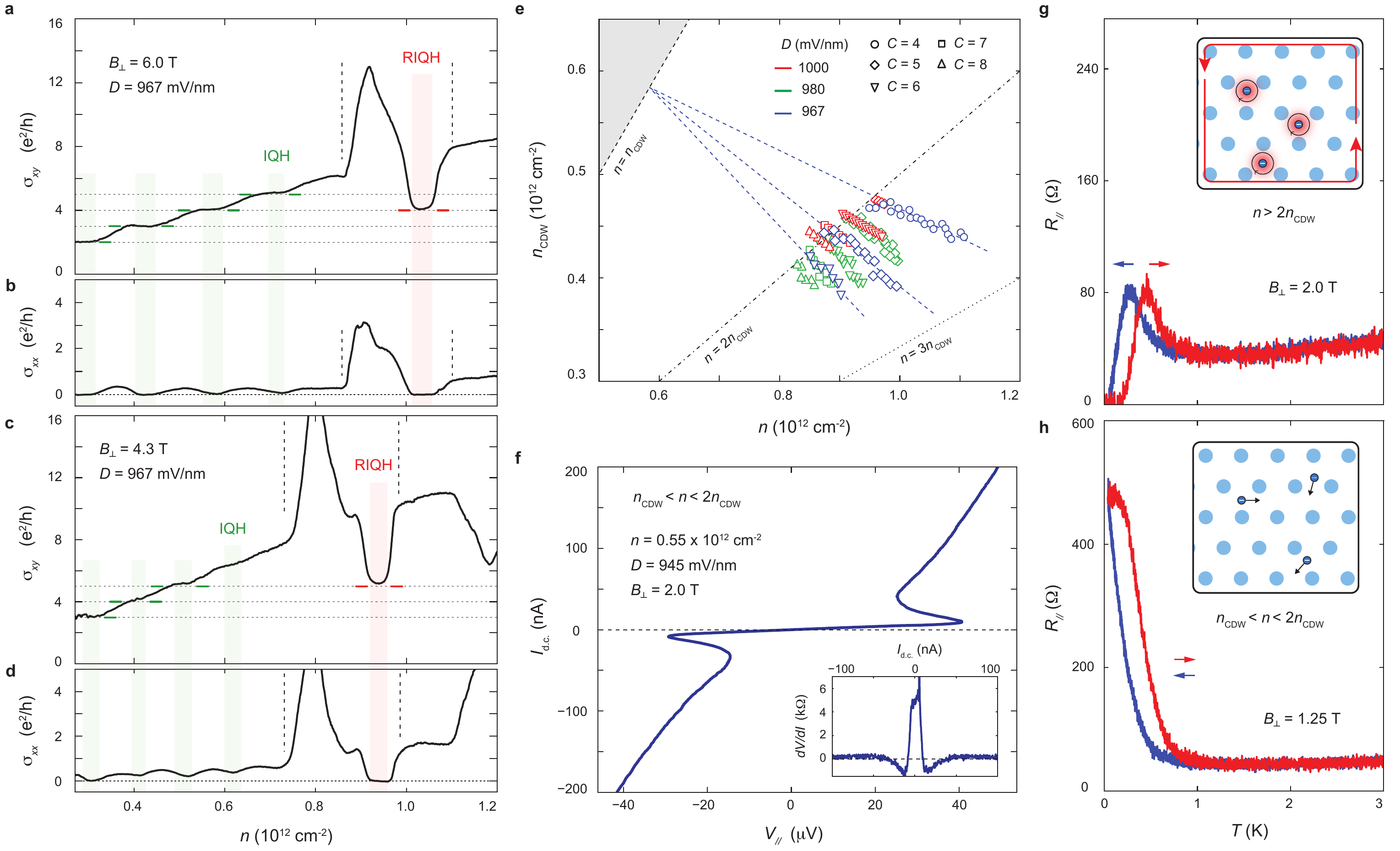}
\caption{
\textbf{Two limits of CDW order.} 
(a--d) Transverse  and longitudinal conductivities, $\sigma_{xy}$ (a, c) and $\sigma_{xx}$ (b, d), as functions of carrier density measured at $D = 967\,\mathrm{mV/nm}$ and perpendicular magnetic fields of (a, b) $B_{\perp}= 6\,\mathrm{T}$,  and  (c, d) $4.3\,\mathrm{T}$, along the horizontal cuts indicated in Fig.~\ref{fig1}e--f. Green and red shaded regions denote conventional quantum Hall states and re-entrant integer quantum Hall states, respectively. 
(e) Number of carriers incorporated into the CDW order, $n_{\mathrm{CDW}}$, as a function of the total carrier density, $n$. The dashed, dash-dotted, and dotted lines denote $n = n_{\mathrm{CDW}}$, $n = 2 n_{\mathrm{CDW}}$, and $n = 3 n_{\mathrm{CDW}}$, respectively. RIQH states are observed only at $2 n_{\mathrm{CDW}} < n < 3 n_{\mathrm{CDW}}$. The gray-shaded region corresponds to $n < n_{\mathrm{CDW}}$, which is physically inaccessible. 
(f) Current--voltage ($I$--$V$) characteristics  as a function of d.c. current bias $I_{\mathrm{d.c.}}$, measured at $n_{\mathrm{CDW}} < n < 2n_{\mathrm{CDW}}$. Bottom inset: differential resistance $dV/dI$ versus $I_{\mathrm{d.c.}}$. The current-driven breakdown of the CDW is evidenced by a pronounced negative dip in $dV/dI$. 
(g, h) Temperature dependence of \Rpara\ measured from the (g) $n > 2n_{\mathrm{CDW}}$ and (h) $n_{\mathrm{CDW}} < n < 2n_{\mathrm{CDW}}$ regions, showing pronounced hysteresis between cooling and warming. Inset: schematic illustration of the CDW orders in two limits. 
}
\label{fig2} 
\end{figure*}

Quantum Hall physics and unconventional superconductivity have evolved for decades as largely separate fields.  The divide stems from the heretofore universal absence of superconductivity in the quantum Hall regime, which is expected because Landau quantization 
breaks time-reversal symmetry and therefore tends to suppress superconducting order. 
Although superconductivity in quantizing magnetic fields is theoretically allowed, 
at least at the mean-field level~\cite{Gruenberg1968QHSC,Rasolt1992QHSC,MacDonald1992QHSC,MacDonald1993QHSC,Song2017QHSC,Chaudhary2021QHSC}, compelling experimental evidence that it survives quantizing fields has been absent. 
Here we show that superconducting and quantum Hall states occur nearly 
simultaneously near the flat-band edge of rhombohedral hexalayer graphene (R6G) ~\cite{Zhou2021RTG,Zhou2021RTG_SC,Zhou2022BLG,Qin2025stripeSC,Han2025chiral,Choi2025SC,Lu2025}
and that both behaviors appear to be shaped by simultaneous CDW order. 

R6G has an unconventional and largely unexplored quantum Hall regime.
In conventional quantum Hall systems, the cyclotron gap, $\Delta_{\mathrm{cyclotron}}$, is
comparable to the Coulomb interaction energy, $\Delta_{\mathrm{Coulomb}}$, 
so that Landau level (LL) mixing is weak and electronic states at fractional filling factors 
are governed primarily by interactions within a single LL~\cite{fukuyama_yoshioka_CDW,yoshioka_lee_CDW}.  
Near the R6G flat-band edge, however, the hierarchy is reversed. Landau quantization creates a dense spectrum of nearly degenerate LLs, illustrated in Fig.~\ref{fig1}a. This unusual LL spectrum is the finite-field cousin of zero field band dispersion that is controlled by competing virtual processes. Processes that connect the opposite R6G surfaces through high-energy intermediate states favor conventional upward curvature, whereas processes that dress the two surfaces separately favor the opposite curvature in the presence of a vertical electric field. The R6G band edge is flattest when these two contributions nearly cancel.
In this case $\Delta_{\mathrm{Coulomb}}\gg\Delta_{\mathrm{cyclotron}}$ and mixing between multiple LLs is 
strong~\cite{allaninfluence1984,inti2013LL,mishra2026CDW}.  %Interactions can hybridize different LL orbitals and fundamentally reconstruct correlated electron states. 
Because of the competing processes, the ground state depends sensitively on the external gate electric field, as well as on carrier-density-dependent screening of that field. Net downward dispersion at small $k$ (Fig.~\ref{fig1}b) leads to unconventional dependence of energy on LL index in which the
$n_{\mathrm{LL}}=0$ orbital LL is pushed to higher energy (Fig.~\ref{fig1}c red solid line), leaving a low-energy sector dominated by LLs with large orbital indices.  This inverted LL hierarchy has important consequences for the interaction physics (see Methods for a more detailed discussion). 

In this work, we investigate the flat-band edge of R6G in a perpendicular magnetic field. We show that Landau quantization near the flat-band edge stabilizes a family of CDW states
whose magneto-transport signatures bear the hallmark of strong LL mixing. 
Most remarkably, CDW order emerges 
intertwined with an unusual superconducting phase that is stable in quantizing magnetic fields, providing the 
first experimental observation of superconductivity in the quantum Hall 
regime~\cite{Gruenberg1968QHSC,Rasolt1992QHSC,MacDonald1992QHSC,Chaudhary2021QHSC}.

The density--electric field map ($n$--$D$) in Fig.~\ref{fig1}d, which plots the longitudinal transport response \Rpara\ at $T = 30$~mK and \Bperp$= 0$ in the regime where the conduction-band edge is fully flattened ($D \sim 1000\,\mathrm{mV/nm}$~\cite{Li2025trashcan,Bernevig2025trashcan}), 
sets the stage for our investigation.   
The map reveals multiple superconducting phases, identified by vanishing resistance and shown in dark blue~\cite{Han2025chiral,Qin2025stripeSC}, labeled SC~i, ii, and iii and outlined by white dashed contours. A few resistive regimes are interspersed between the superconducting phases, forming a distinctive checkerboard-like pattern across the flat-band line in the phase diagram. We label these regions I, II, and III in order of increasing carrier density, as indicated by black dashed contours in Fig.~\ref{fig1}d.
The line in the $n$--$D$ map along which these states are clustered is close to the line of optimal band flatness.

The rich non-zero \Bperp\ phase space is governed by three independent parameters: $n$, $D$, and \Bperp. This multidimensional phase space can be explored through two-dimensional cuts, including $n$--\Bperp\ planes at fixed $D$ and $n$--$D$ planes at fixed \Bperp. 
Figs.~\ref{fig1}e--f plot the longitudinal and transverse resistances, \Rpara\ and \Rperp, as functions of $n$ and \Bperp, measured at $D = 967\,\mathrm{mV/nm}$ (white arrow in Fig.~\ref{fig1}d). On the low-density side of the $n$--\Bperp\ map, a series of quantum oscillations forms a characteristic fishnet-like pattern (see also Fig.~\ref{nB_SC}). These oscillations occur when unconventional conduction band 
dispersion is dominant, which leads to a band-minimum at finite momentum and an associated annular Fermi surface with coexisting electron- and hole-like carriers.
Portions of this map connect continuously to the zero-field regimes I--III and are labeled accordingly. Weak-field quantum oscillations are notably absent within these regimes, signaling flatter-band edges that strongly suppress LL separation. The flat-band regimes disappear at very large \Bperp when 
the LL degeneracy becomes comparable to the density-of-states integrated across the flat-band edge.

We focus first on regime II which is most unconventional.  Its edges are delineated by resistive peaks, highlighted by black dashed lines. This regime exhibits the rich interplay between CDW order, quantum Hall effects, and superconductivity that is the primary focus of this work. In the following, we first characterize its CDW order before examining the coexisting superconducting phase.  

CDW order is most strongly manifested on the high-density side of regime II, where it
is revealed by a sequence of integer quantum Hall states (Fig.~\ref{fig1}e--f).    
The underlying CDW order is identified in Figs.~\ref{fig2}a--d, which plot the $n$ dependence of the longitudinal and Hall conductivities, $\sigma_{xx}$ and $\sigma_{xy}$, extracted from horizontal line cuts of Fig.~\ref{fig1}e--f along the red and blue bars. A sequence of integer quantum Hall (IQH) states at low density, marked by green-shaded stripes, highlights the $n$-dependence expected from the conventional quantum Hall regime. Within regime II, however, $\sigma_{xy}$ deviates from its monotonic density dependence and forms a re-entrant Hall plateau. Combined with vanishing $\sigma_{xx}$, these behaviors establish the hallmark signatures of the quantum Hall effect.  
The states develop along well-defined trajectories in the $n$--\Bperp\ map.  Since the Hall conductance quantum number of these states, which equals the total Chern number of occupied LLs, is smaller than the filling factor, we conclude that the LLs must be dressed by CDW order that yields several 
occupied LLs with vanishing Chern numbers.  This behavior is consistent with the classical electron behavior --- lattice formation and localized electron bands with vanishing Chern numbers --- expected when kinetic energy is entirely absent.  If we assume that the Hall conductivity is reduced because some electrons have localized near lattice sites, we can read the unit cell area 
$A_{\mathrm{CDW}} \equiv n_{\mathrm{CDW}}^{-1}$ of the density wave from the measured Hall conductance, $n_{\mathrm{CDW}} =n-\sigma_{xy}B /((e^2/h) \Phi_0)$.
Here $n_{\mathrm{CDW}}$, the density that can be accommodated by one CDW induced band in the absence of a magnetic field, is identified as the density of localized electrons in the CDW state.

The reentrant states we observe differ qualitatively from those observed previously in semiconductor quantum wells and in monolayer graphene, which 
are associated with localization of the charged excitations of an undressed Landau level
and yield a Hall quantization index equal to the integer closest to the filling factor ~\cite{Eisenstein2002bubble,Fogler2002stripe,Goerbig2003RIQH,Goerbig2004electronsolid,Xia2004bubble,Deng2012reentrant,Gervais_3rdLL,Liu2012reentrant,Halperin2020FQHE,Chen2019RIQH}.
The re-entrant integer quantum Hall (RIQH) states in Fig.~\ref{fig2}a--d, however, exhibit Hall quantization that is substantially smaller than the total LL filling. For example, the RIQH state in Fig.~\ref{fig2}a--b appears at an effective LL filling of approximately $\nu \sim 8$, yet the measured Hall quantization is only $\sigma_{xy}=4e^2/h$.
This behavior is a natural consequence of CDW order arising from the reconstruction of a manifold of multiple nearly degenerate LLs. As illustrated schematically in Fig.~\ref{LL}b, the CDW state can
be viewed as one in which some occupied Landau levels have reorganized to form a Wigner crystal, while higher index Landau levels retain their nonzero Chern numbers, which likely form broad bands of unoccupied states localized in the interstitial regions of nearby Wigner crystal states,
and continue to contribute to the quantized Hall response.  Our observations 
therefore establish the 
realization of a unique type of topological CDW order, a subject of broad recent interest~\cite{Seiler2022BLG,Tan2024hallcrystal,Dong2024hallcrystal,Soejima2024hallcrystal,Dong2024AHC,Patri2024QAH,Dong2024RPG,Lu2025}.

Figure~\ref{fig2}e plots $n_{\mathrm{CDW}}$ as a function of total carrier density $n$ for a series of RIQH states measured across $n$--$B_{\perp}$ maps at different values of $D$ (see Fig.~\ref{CDWanalysis}).  These results show that 
the density of localized electrons increases slowly with 
gate electric field, as the flat-band region in momentum space slowly expands, and decreases 
with density as the gate-field is screened (also see Fig.~\ref{nCDW}a--b).
The observation that $n_{\mathrm{CDW}}$ is not fixed implies that the spatial periodicity of the CDW is continuously tunable. This behavior contrasts sharply with that of conventional charge-ordered phases, whose periodicity is typically commensurate with an underlying crystal lattice~\cite{gruner2018density,monceau2012electronic}.  
Our observations point to interaction-driven charge order in which some electrons are reorganized into localized bands, while others take advantage of the possibility of opening gaps between Landau levels to yield a CDW period that is both electron-density and magnetic-field dependent. 
In Fig.~\ref{fig2}e, the dashed lines serve as guides to the eye for the evolution of $n_{\mathrm{CDW}}$ with $n$ (see also Fig.~\ref{nCDW}c--d). We note that for all fixed values of D, the data extrapolate toward a common point satisfying $n_{\mathrm{CDW}} = n$, {\em i.e.} to Wigner crystal states.  The extrapolated value coincides with the left boundary of regime II, 
suggesting that, because of the limited extent of the flat-band regime,
this is the maximum density at which it is possible to construct Wigner crystal states.
Upon further doping, most additional carriers must be itinerant, although $n_{\mathrm{CDW}}$ does evolve throughout the phase diagram (Fig.~\ref{nCDW}). As a direct consequence of the variation of
$n_{\mathrm{CDW}}$, the slopes of the RIQH states in the $n-$\Bperp\ plane show pronounced deviations (Fig.~\ref{mismatch}) from Streda formula expectations ~\cite{Streda1983streda,Huang2025streda} that would apply if the CDW period was fixed. 

\begin{figure*}
\includegraphics[width=1\linewidth]{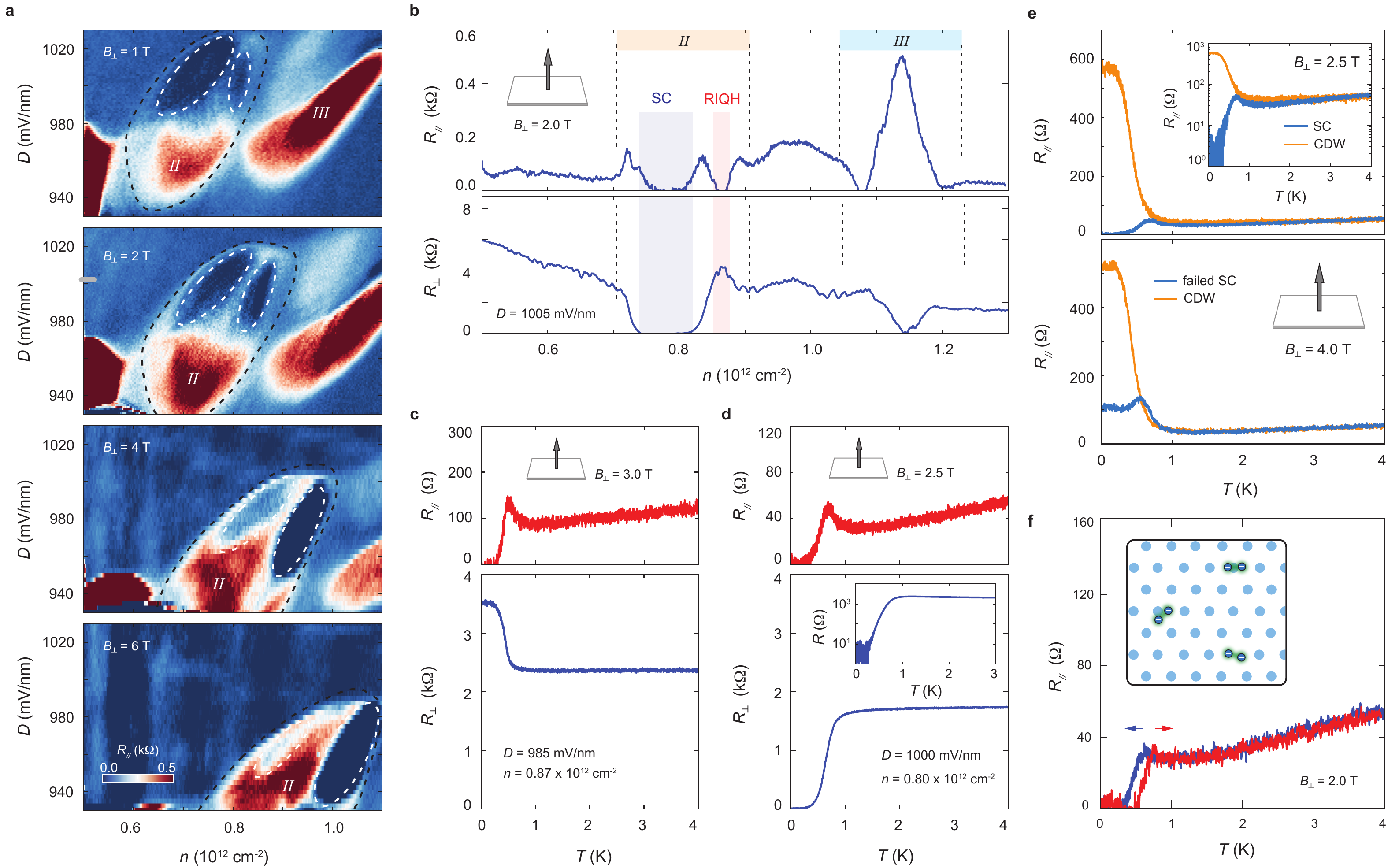}
\caption{\textbf{Superconductivity emerging from LL formation.}  
(a) $n$--$D$ maps of \Rpara\ measured at fixed perpendicular magnetic fields $B_{\perp}$. 
(b) Line cut of the longitudinal (top) and transverse (bottom) resistance as a function of $n$, measured at $D = 990\,\mathrm{mV/nm}$ and $B_{\perp} = 2\,\mathrm{T}$, corresponding to the horizontal gray bar in panel (a). Vertical dashed lines mark the density ranges of regimes II and III. Within regime II, two distinct low-temperature phases emerge from the CDW background: a RIQH state, and superconductivity (SC), marked by vanishing longitudinal resistance. 
(c, d) Temperature dependence of the longitudinal (top) and transverse (bottom) resistance measured in (c) the RIQH state and (d) the superconducting state.  The inset in the bottom (d) panel shows the same \Rperp--$T$ trace on a logarithmic vertical scale, highlighting the vanishing resistance in the superconducting phase.
(e) Temperature dependence of \Rpara\ measured in the superconducting phase (blue traces) and in the CDW order at $2 > n/n_{CDW} > 1$ (orange traces), measured at \Bperp$=2.5$ T (top) and $4.0$ T (bottom). The inset in the top panel shows the same \Rpara--$T$ trace on a logarithmic vertical scale, highlighting the vanishing resistance in the superconducting phase.
(f) Temperature dependence of \Rpara\ measured from the superconducting phase, showing pronounced hysteresis between cooling and warming. Inset: schematic illustration of Cooper pairing instability, emerging from a background of CDW order. 
}
\label{fig3}
\end{figure*}

\begin{figure*}
\includegraphics[width=0.88\linewidth]{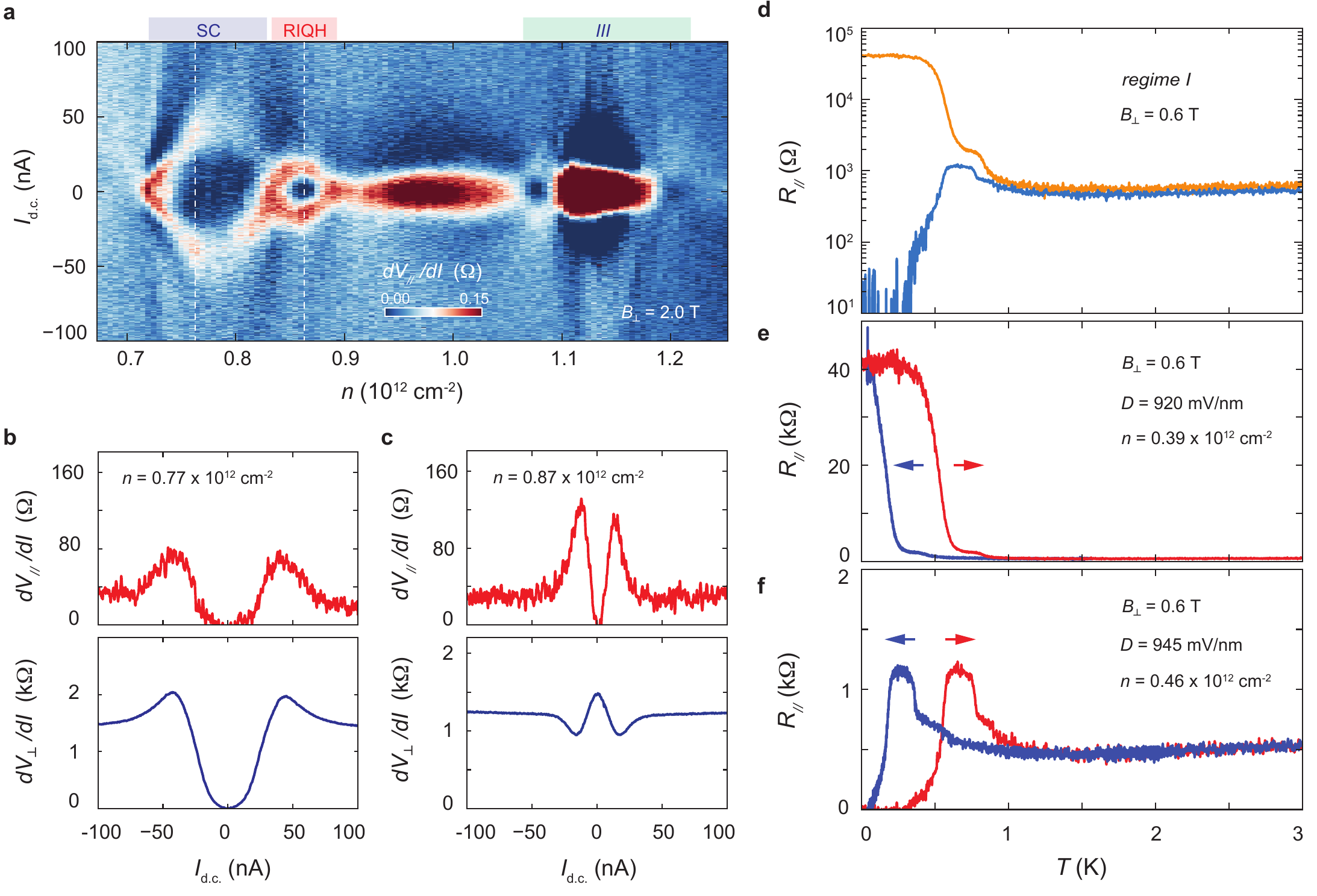}
\caption{\textbf{$I$--$V$ characteristics.} 
(a) Longitudinal differential resistance, $dV_{\parallel}/dI$, as a function of $I_{\mathrm{d.c.}}$ and $n$ across regimes II and III, measured at $B_{\perp}=2\,\mathrm{T}$ and $D = 1005$~mV/nm. 
(b, c) Differential resistance measured as a function of $I_{\mathrm{d.c.}}$ in the longitudinal (top) and transverse (bottom) channels at the (b) superconducting and (c) RIQH states, respectively, with locations indicated by the white dashed lines in panel (a). 
%(d) $R_{\parallel}$ as a function of $n$ and $D$ across resistive regime I, measured at $B_{\perp}=0.6\,\mathrm{T}$. 
(d) Temperature dependence of \Rpara\ measured in the CDW (orange) and superconducting (blue) states in regime I. 
(e, f) \Rpara--$T$ traces measured during cooling (blue) and warming (red) for the same (e) CDW and (f) superconducting states shown in panel (d). 
}
\label{fig4} 
\end{figure*}

%The property that Hall quantization emerges only at $n > 2 n_{\mathrm{CDW}}$ \blue{offers an intriguing observation.}
%may suggest that the CDW order has a honeycomb spatial arrangement which generates two flat bands, and more dispersive higher energy bands with larger LL separations.  

While Hall quantization is absent when $1 < n/n_{\mathrm{CDW}} < 2$, CDW order in this regime is indicated by the distinctive non-linear current--voltage ($I$--$V$) characteristics shown in Fig.~\ref{fig2}f. Near zero d.c. bias ($I_{\mathrm{d.c.}}=0$), the differential resistance $dV/dI$ is consistent with a resistive state. As $I_{\mathrm{d.c.}}$ increases, however, the system undergoes a sharp transition into a highly conductive regime at large bias (inset of Fig.~\ref{fig2}f). This transition, occurring near $I_{\mathrm{d.c.}}=\pm10$~nA, is accompanied by pronounced negative dips in $dV/dI$.
Such $I$--$V$ characteristics are a well-established hallmark of Wigner solids in semiconductor quantum wells. In particular, the negative dip in $dV/dI$ reflects a current-driven breakdown of the Wigner solid~\cite{Csathy2007CDW}. The temperature dependence of the $I$--$V$ response reveals a sharply defined transition near $T = 0.5$~K, above which the nonlinear response disappears and the $I$--$V$ characteristic becomes fully ohmic (see Fig.~\ref{IVT}).  Together, these findings provide compelling evidence for CDW order throughout region II. 

The presence of CDW order is further supported by the nature of the thermal melting transitions. Fig.~\ref{fig2}g--h show \Rpara$-T$ traces measured on either side of the boundary $n = 2 n_{\mathrm{CDW}}$. In both limits, pronounced hysteresis is observed between warming and cooling measurements. This hysteresis is characteristic of a first-order melting transition with superheating and supercooling, providing thermodynamic evidence for a crystalline phase ~\cite{Reddy2026melting,Chen2006melting,Xiang2025melting,Ma2020melting}.

%{\color{red} {\bf Should we drop this.  I am uncomfortable calling it a Wigner solid because it is not very resistive and because that word is reserved for $n=n_{CDW}$.} How should we adjust the figures? Overall, our observations reveal two distinct forms of CDW order. CDW-I emerges in coexistence with fully occupied LLs and is manifested through the RIQH effect, as illustrated schematically in the inset of Fig.~\ref{fig2}g. In contrast, CDW-II closely resembles the Wigner-solid phases observed in conventional quantum Hall systems,  as illustrated in the inset of Fig.~\ref{fig2}h. As summarized schematically in Fig.~\ref{LL}b, both CDW orders can be understood as natural consequences of interaction-driven reconstruction within a manifold of strongly mixed LLs near a flat-band edge.}

One of the most intriguing aspects of the CDW order we observe is its coexistence with superconductivity when $1 < n/n_{\mathrm{CDW}} < 2$. Fig.~\ref{fig3}a plots $n$--$D$ maps of \Rpara\ measured around regime II at different values of \Bperp. In the $n$--$D$ plane, the boundary of regime II is delineated by the onset of finite \Rpara, outlined by black dashed contours. With increasing \Bperp, this regime shifts toward higher $n$ and lower $D$, consistent with the trend observed in the $n$--\Bperp\ maps shown in Fig.~\ref{fig1}e--f. 
Although most of regime II remains resistive, two pockets of vanishing \Rpara\ emerge
at finite \Bperp, highlighted by white dashed contours in Fig.~\ref{fig3}a. The higher-density pocket corresponds to the RIQH effect discussed previously. As we explain below, 
the lower-density pocket corresponds to a superconducting phase stabilized by a perpendicular magnetic field. In the $n$--$B_{\perp}$ map, this superconducting phase forms an extended zero-resistance region, as shown in Fig.~\ref{nB_SC}.

Fig.~\ref{fig3}b plots the longitudinal and transverse resistances, \Rpara\ and \Rperp, as functions of $n$ at \Bperp$=2$~T, measured along the horizontal white arrow in Fig.~\ref{fig3}a and intersecting both RIQH and superconducting phases. The RIQH state exhibits vanishing \Rpara\ and quantized \Rperp, highlighted by the red-shaded vertical stripe. In contrast, the superconducting phase exhibits vanishing resistance in both \Rpara\ and \Rperp, marked by the blue-shaded vertical stripe. This transport response points to a diverging conductivity tensor consistent with a superconducting phase.

Upon varying temperature, both RIQH and superconducting phases exhibit sharply defined transitions. At the onset of the RIQH state, the vanishing of \Rpara\ is accompanied by a pronounced enhancement of \Rperp\ (Fig.~\ref{fig3}c). This indicates that the number of mobile charge carriers participating in transport is reduced below the transition temperature, consistent with the formation of CDW order. By contrast, the low-density state is characterized by the simultaneous suppression of both \Rpara\ and \Rperp\ (Fig.~\ref{fig3}d). In the presence of a large perpendicular magnetic field, few mechanisms can account for the concurrent vanishing of longitudinal and transverse resistance, providing a stringent diagnostic. Moreover, as shown in Fig.~\ref{RT_SC}, the measured resistance in this state remains within the experimental noise floor around \Rpara$=0$. We therefore attribute it to a superconducting phase, in which dissipationless bulk transport effectively shorts the Hall voltage across the sample.

We note that $n/n_{\mathrm{CDW}} = 2$ defines a boundary separating the RIQH and superconducting phases. This strongly suggests that the emergence of CDW order reconstructs the electronic band structure (see Fig.~\ref{nB_schematic}c). Superconductivity is stabilized when itinerant electrons occupy the first reconstructed sub-band ($n<2n_{\mathrm{CDW}}$), whereas Landau quantization emerges once carriers begin to populate the second sub-band ($n>2n_{\mathrm{CDW}}$). The emergence of Hall quantization only near $n = 2 n_{\mathrm{CDW}}$
is intriguing, and might suggest that the gaps between itinerant Landau levels are larger when the area per Landau level is commensurate with the area per CDW period.  Together, these observations point to the unusual quantum phases arising from CDW order with spontaneous self-doping ~\cite{Dong2026crystal,Feng2026self}.

The emergence of superconductivity raises an important question: how can LL formation give rise to a Cooper-pairing instability? While the microscopic mechanism remains unclear, we show below that the superconducting phase develops on a background of CDW order, pointing to an interplay between these two phases as a key ingredient underlying the superconducting stability. This interplay is reflected by a particular feature in the \Rpara--$T$ traces in Fig.~\ref{fig3}d, where a pronounced peak appears in \Rpara\ immediately above the superconducting transition temperature. In Fig.~\ref{fig3}e, we explore the origin of this peak by comparing \Rpara--$T$ traces measured in the superconducting phase (blue traces) with those measured in the normal state (orange traces). Not only does the emergence of superconductivity coincide with the onset of the CDW order, but the resistance peak above the superconducting transition also closely matches the resistive onset associated with the Wigner-solid-like behavior (top panel of Fig.~\ref{fig3}e). In particular, when superconductivity is partially suppressed at \Bperp\ $=4\,\mathrm{T}$ (bottom panel of Fig.~\ref{fig3}e), the corresponding \Rpara--$T$ curve develops an upturn that closely follows the resistive behavior in the normal state. These observations suggest that upon cooling, the CDW order is the first to emerge. 

Furthermore, the superconducting transition exhibits thermal hysteresis that is reflective of the CDW order (Fig.~\ref{fig3}f). Conventionally, the transition from the superconducting to the normal state is expected to be second order. We therefore argue that the observed hysteresis instead reflects a first-order melting transition of the coexisting CDW order. In this picture, it is CDW order 
in the parent state that stabilizes the formation of Cooper pairs and their subsequent condensation.

The connection between superconductivity and CDW order is also reflected in the $I$--$V$ characteristics. Fig.~\ref{fig4}a plots the differential longitudinal resistance $dV_{\parallel}/dI$ as a function of $I_{\mathrm{d.c.}}$ and $n$ across regimes II and III, capturing the nonlinear transport behavior of both the superconducting and RIQH states. The superconducting state exhibits a current-driven transition into a normal metallic regime, with the peak in differential resistance defining a critical current of $I_{\mathrm{d.c.}} \approx 40$~nA (Fig.~\ref{fig4}b). The RIQH state displays a qualitatively similar evolution: a current-driven transition marked by a peak in $dV/dI$, accompanied by a suppression in the differential Hall response (Fig.~\ref{fig4}c). 

As shown in Fig.~\ref{fig4}a, pronounced negative dips in $dV/dI$ are observed immediately above the critical current of both the superconducting and RIQH states (see also Fig.~\ref{IVn}). This indicates that with increasing $I_{\mathrm{d.c.}}$, the superconducting and RIQH states are suppressed first, while a further increase in $I_{\mathrm{d.c.}}$ induces a second transition, resembling the current-driven breakdown of a Wigner-solid-like state.   

Despite its intimate interplay with CDW order, the onset of superconductivity is well described as a Berezinskii--Kosterlitz--Thouless (BKT) transition. This is established by extracting the power-law exponent $\beta$ from the $I$--$V$ characteristics ($V \propto I^{\beta}$). As shown in Fig.~\ref{BKT}, $\beta$ exceeds three at low temperature and evolves abruptly toward unity with increasing temperature, consistent with the emergence of quasi-long-range phase coherence in the superconducting phase~\cite{Merchant2001crossover}. This behavior defines the BKT transition temperature $T_{\mathrm{BKT}}$, which is in excellent agreement with the superconducting transition inferred from the \Rpara--$T$ traces in Fig.~\ref{fig3}d (see also Fig.~\ref{onset}).

It should be noted that the non-linear transport associated with CDW order is universally observed across regimes I--III. Fig.~\ref{CDW} shows nonlinear $I$--$V$ characteristics with sharply defined current-driven breakdown throughout all three resistive regimes, in sharp contrast to the linear $I$--$V$ response observed outside these regions (Fig.~\ref{linearIV}). In addition, Fig.~\ref{fig4}d shows the \Rpara--$T$ trace measured in regime I (orange trace; see also Fig.~\ref{linecut}), which exhibits a sharp onset near $0.5$~K. This onset exhibits pronounced hysteresis between cooling and warming measurements (Fig.~\ref{fig4}e), consistent with a first-order melting transition of the underlying CDW order.

While the thermal melting transition of the CDW order consistently occurs above $0.5$~K, 
the current-driven breakdown threshold varies substantially with $n$ and $D$, ranging from $2$--$3$~nA to $20$--$30$~nA (see Fig.~\ref{CDW}). This rules out a simple Joule-heating effect as the origin of the nonlinear $I$--$V$ response. Furthermore, the nonlinear $I$--$V$ curve disappears above the thermal melting transition, establishing a direct connection between the two phenomena. This allows us to rule out alternative mechanisms for nonlinear transport, such as Schwinger-Landau-Zener breakdown ~\cite{Oka2003breakdown,Oka2005breakdown}.

Unlike a conventional Wigner solid, which behaves as a true insulator, the Wigner-solid-like states in regimes I--III remain relatively conductive, which is evidenced by the slope in the $I$--$V$ curve near zero bias. This behavior is consistent with the coexistence between localized and itinerant electrons ~\cite{Dong2026crystal,Feng2026self}. 
Regime I is, however, most resistive and has transport characteristics that are consistent with a small value of $n-n_{CDW}$.  Signatures of superconductivity are also observed in the resistive regimes I and III, where they onset together with coexisting CDW order. For instance, near the low- and high-density edges of regime III, Fig.~\ref{fig4}a reveals two additional states with vanishing $dV/dI$ near zero bias. These features likely correspond to incipient superconducting and quantum Hall states, analogous to those observed in regime II.

A similar, but more robust, superconducting phase is observed in regime I, appearing as a zero-resistance state adjacent to the CDW order (see Fig.~\ref{linecut} and Fig.~\ref{nB_SF2}). Upon cooling, superconductivity emerges simultaneously with the CDW order, as shown by the blue trace in Fig.~\ref{fig4}d. Moreover, the superconducting transition exhibits pronounced thermal hysteresis (Fig.~\ref{fig4}f). These observations closely parallel those in regime II (Fig.~\ref{fig3}e, f), further supporting an intimate connection between superconductivity and CDW. In both regimes, the apparent superconducting transition is bounded by the thermal melting of the underlying CDW order, above which superconductivity is no longer stable.

Taken together, our observations establish a new regime of quantum Hall physics that directly connects charge ordering and superconductivity within the same manifold of strongly mixed LLs. The realization of superconductivity in the quantum Hall regime suggests that quasiparticle interactions with low-energy collective excitations of the finite-B CDW-state provide more favorable retarded interactions that overcome the negative influence of broken time reversal.  While the  microscopic origin of the superconducting phase, as well as its precise interplay with CDW order, remains an open question, this new regime is likely to motivate broad future efforts aimed at understanding the emergence of CDW and superconductivity from strongly mixed LLs.

\bibliography{Li_ref}

\newpage
\clearpage

\section*{Methods}

\renewcommand{\thefigure}{M\arabic{figure}}
\renewcommand{\theHfigure}{M\arabic{figure}}
\renewcommand{\theequation}{M\arabic{equation}}
\renewcommand{\theHequation}{M\arabic{equation}}
\renewcommand{\thetable}{M\Roman{table}}
\renewcommand{\theHtable}{M\Roman{table}}

\setcounter{figure}{0}
\setcounter{equation}{0}
\setcounter{table}{0}

%\makeatletter
%\cref@old@addtoreset{figure}{}
%\cref@old@addtoreset{equation}{}
%\cref@old@addtoreset{table}{}
%\makeatother

In this section, we provide additional analysis to further substantiate the results presented in the main text, with expanded discussion of the transport signatures of CDW order and its hierarchical relationship to the RIQH and superconducting phases.

\begin{figure*}
\includegraphics[width=0.95\linewidth]{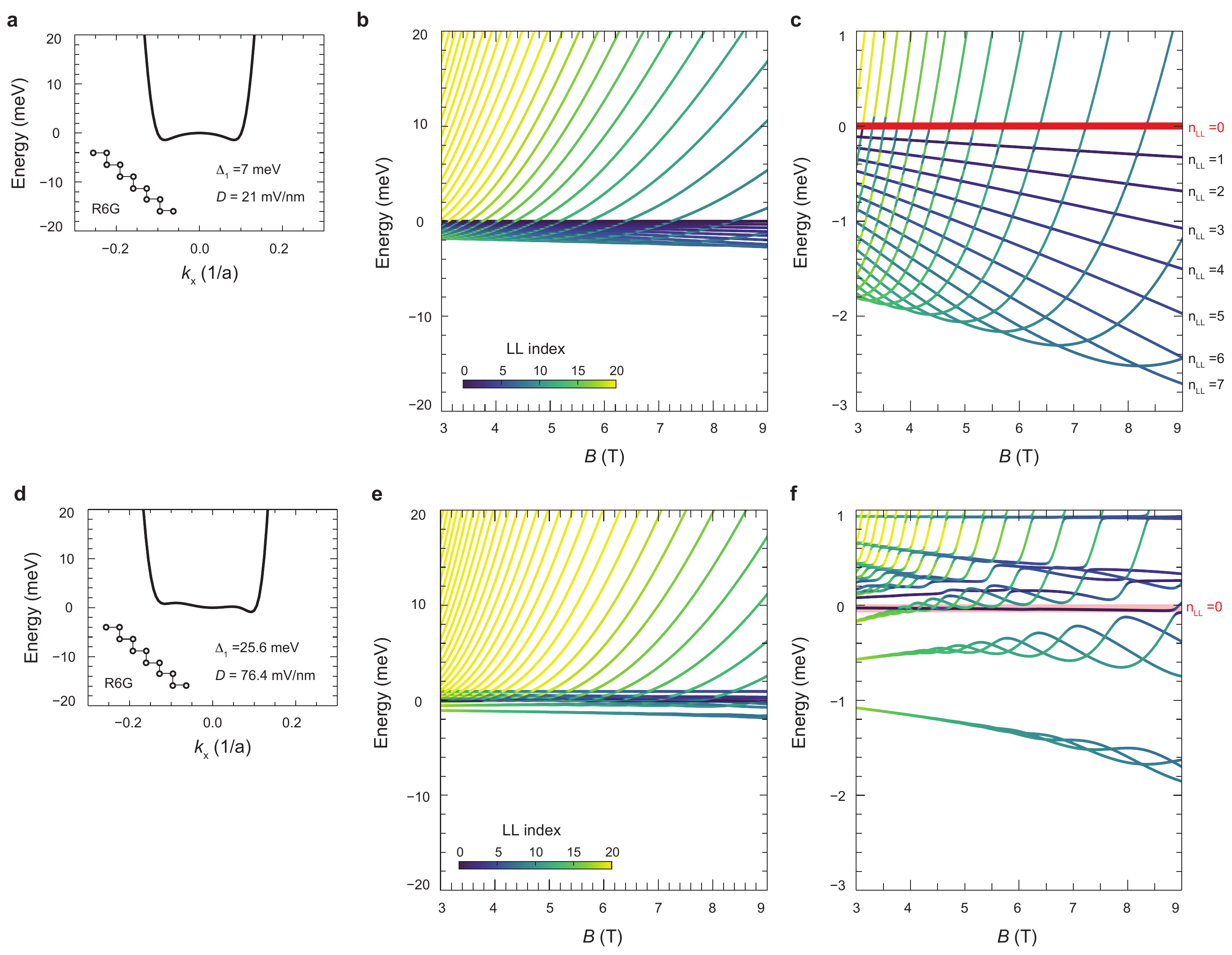}
\caption{
\textbf{Landau quantization near a RG flat-band edge.}
(a) Conduction-band dispersion of R6G calculated using the Slonczewski--Weiss--McClure (SWMc) parameters $\gamma_0$ and $\gamma_1$.
(b,c) Landau-level spectrum derived from the band structure in (a). Panel (c) provides a zoomed-in view of the low-energy edge of the spectrum. The red solid line indicates the $n_{\mathrm{LL}} = 0$ LL and line colors
are used to indicate the orbital LL index associated with a level.  
(d) Conduction-band dispersion of R6G calculated including higher-order SWMc parameters $\gamma_3$, $\gamma_2$, and $\delta$. 
(e,f) Landau-level spectrum derived from the band structure in (d). Panel (f) provides a zoomed-in view of the low-energy edge of the spectrum. The model and parameters are taken from Ref.~\cite{koshinoABC_model,zhangDFT_ABC_trilayer}. Lines are shaded to indicate the 
expectation value of the orbital index operator of a level.  
The red shaded region indicate high weight of the $n_{\mathrm{LL}} = 0$ LL.
}
\label{band}
\end{figure*}

\begin{figure*}
\includegraphics[width=1\linewidth]{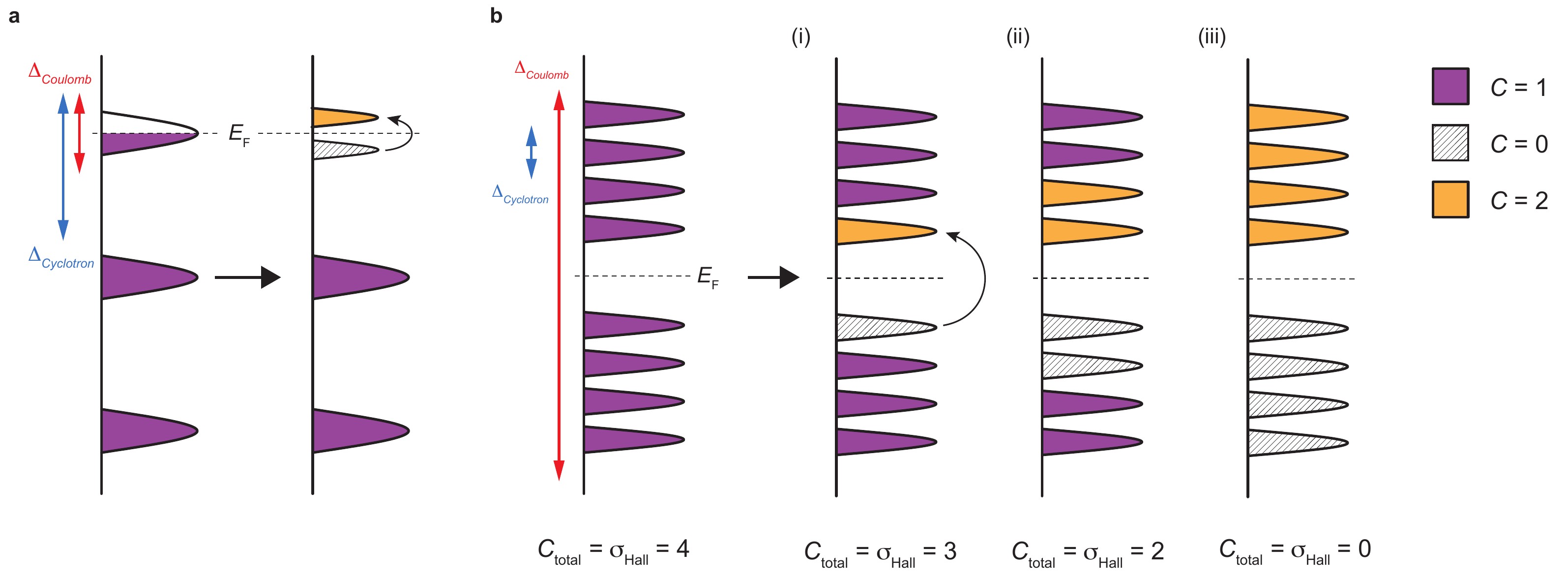}
\caption{
\textbf{CDW order formation and LL Chern number.} 
Schematic diagram illustrating the emergence of CDW order through changes in the Chern numbers of occupied LLs. The color scale represents the Chern number $C$ associated with a LL. Purple denotes $C = 1$, corresponding to the Landau levels of the single-particle Hamiltonian. 
Hashed regions denote $C = 0$, arising from a Wigner-solid-like charge density wave (CDW) order. Orange denotes $C = 2$, which emerges from strong Landau-level mixing.  
Level crossings occur when charge order mixes Landau levels, preserving the total Chern number. 
(a) In conventional quantum Hall systems, the Coulomb interaction energy, $\Delta_{\mathrm{Coulomb}}$ (red line), is smaller than the Landau-level separation, $\Delta_{\mathrm{Cyclotron}}$ (blue line). In this limit, the formation of CDW order splits the LL into two subbands, one occupied and one empty. Owing to the CDW order, electrons in the 
occupied subband are localized with Chern number $C=0$, whereas CDW order with localized holes leads to occupied subbands with $C=1$.  
(b) Near a flat-band edge, $\Delta_{\mathrm{Coulomb}}$ exceeds $\Delta_{\mathrm{Cyclotron}}$, generating strong LL mixing. This schematic illustrates different possible outcomes of level mixing when the system has filling factor $\nu=4$ and contains the equivalent filling of eight LLs in the low energy range. When charge order is weak all LLs retain $C = 1$, resulting in a Hall conductivity of $4e^2/h$. (i-ii) Low many-body energies are achieved when LLs are mixed to occupy orbitals that are localized on lattice sites to reduce the self-interaction cost of CDW formation.  Because they are built from localized
orbitals the new levels have $C = 0$ and pass their native Chern number to higher energy LLs when crossing events occur.  The occupied Landau levels have total Chern numbers that are non-zero but smaller than the filling factor.  Some occupied Landau levels must have Chern numbers larger than $C=1$ due to LL mixing
because of the conservation of total Chern number.  iii) Wigner crystal states must have $C=0$ for 
all occupied Landau levels.
}
\label{LL} 
\end{figure*}

\begin{figure*} \includegraphics[width=0.82\linewidth]{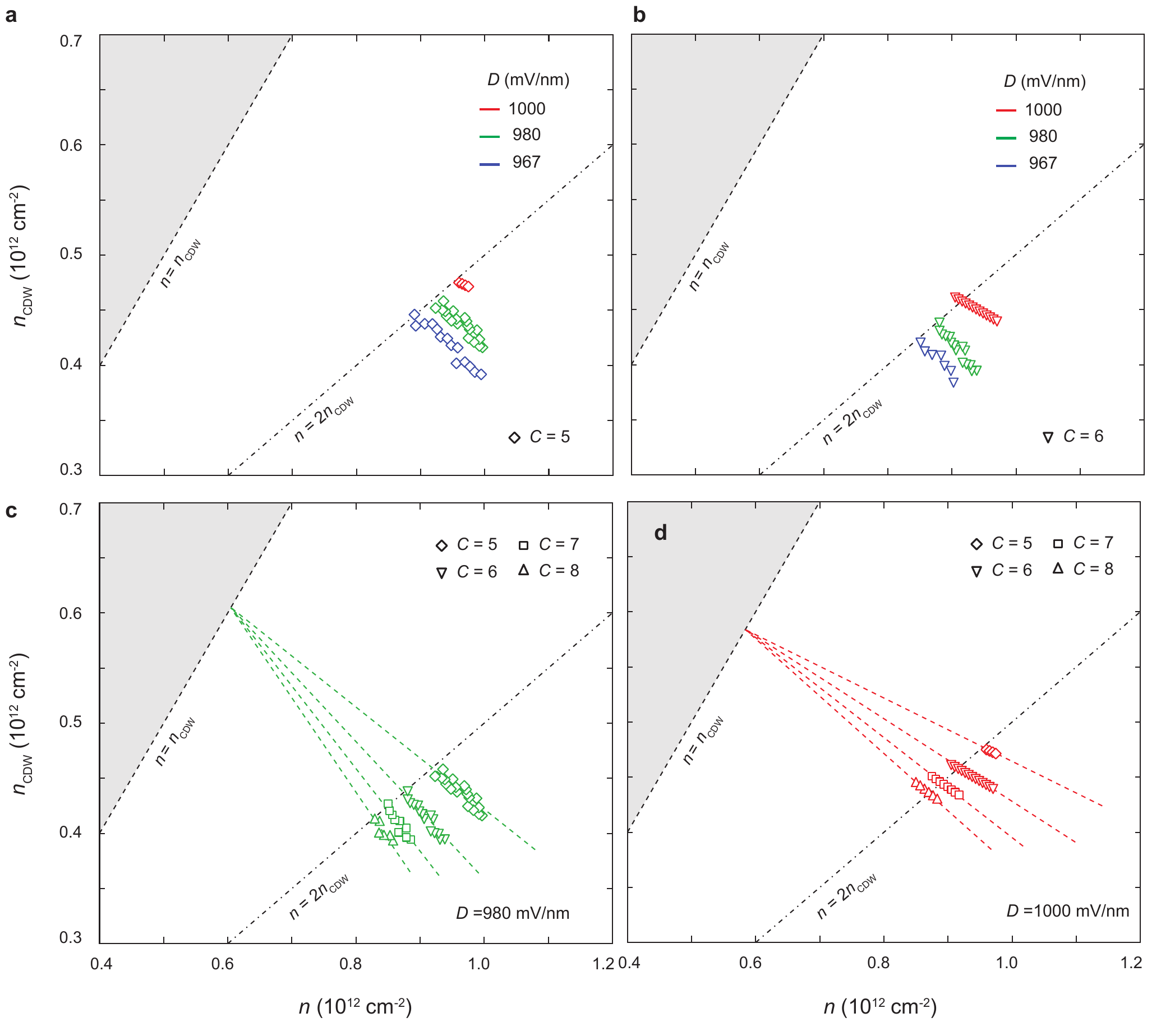} 
\caption{\textbf{Evolution of $n_{\mathrm{CDW}}$.} 
$n_{\mathrm{CDW}}$ extracted from the RIQH states as a function of total carrier density, $n$. 
(a,b) RIQH states with Chern number (a) $C=5$ and (b) $C=6$, measured at different displacement fields, $D$. At a fixed $n$, RIQH states with the same Chern number exhibit larger $n_{\mathrm{CDW}}$ with increasing $D$. 
(c,d) $n_{\mathrm{CDW}}$ extracted from RIQH states measured at (c) $D=980$ mV/nm and (d) $D=1000$ mV/nm. Symbols denote states with different Chern numbers. For each value of $D$, the data extrapolate toward a common point satisfying $n_{\mathrm{CDW}}=n$, corresponding to the limit of vanishing Landau-level density, $n_{\mathrm{LL}}\rightarrow 0$, at $B_{\perp}=0$. The dashed (dash-dotted) line denotes $n_{\mathrm{CDW}}=n$ ($n_{\mathrm{CDW}}=0.5\,n$). The gray-shaded region corresponds to $n_{\mathrm{CDW}}>n$, which is physically inaccessible. Intriguingly, $n_{\mathrm{CDW}} = 0.5\,n$ appears to define a boundary for RIQH stability, with all observed RIQH states occurring in the regime $n_{\mathrm{CDW}} < 0.5\,n$. This corresponds to a situation in which fewer than half of the carriers participate in the CDW order, while the majority of carriers remain itinerant.
} 
\label{nCDW} 
\end{figure*}

\begin{figure*}
\includegraphics[width=0.99\linewidth]{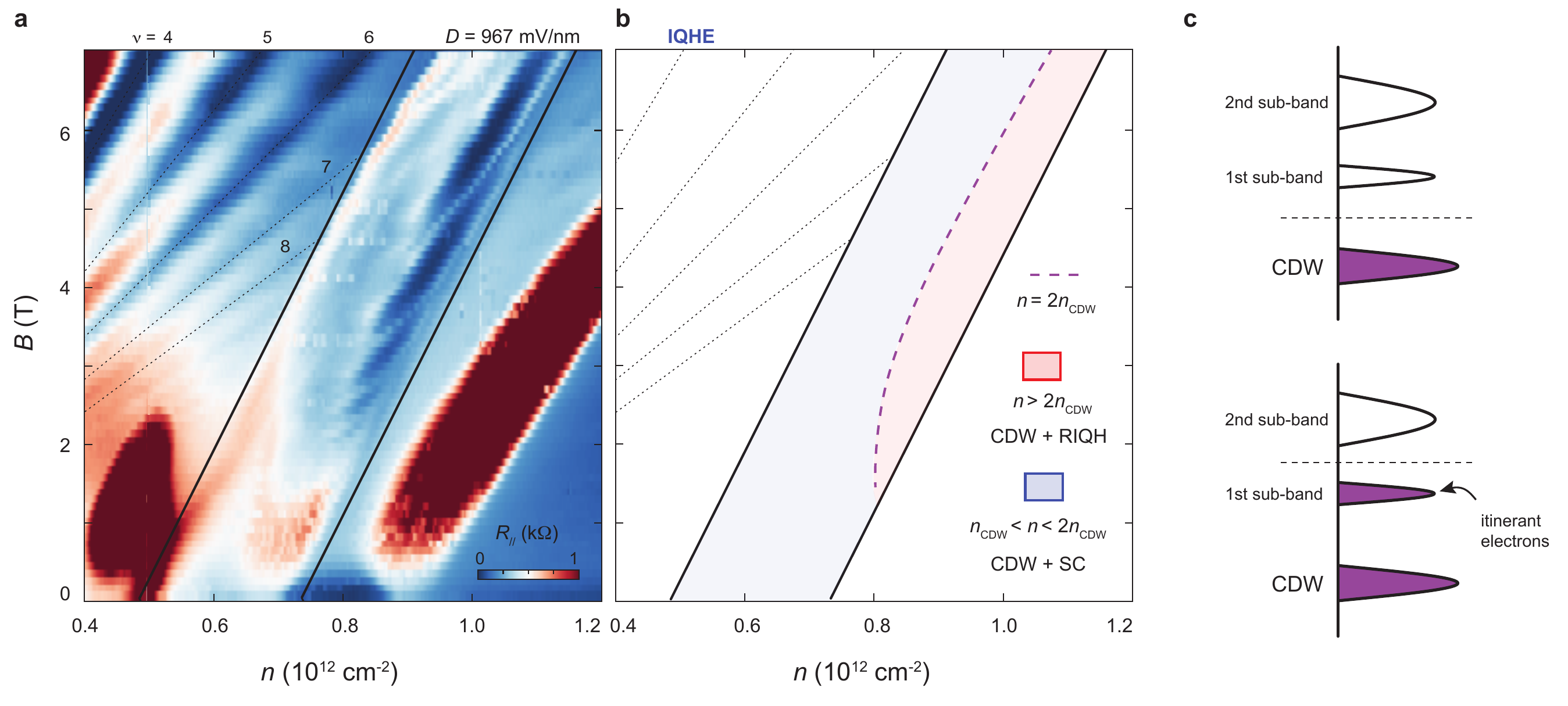}
\caption{\label{nB_schematic}
\textbf{Filling reconstructed sub-bands.}
(a) $n$--\Bperp\ map of \Rpara\ measured across regime II at $D = 967$ mV/nm. Black dashed lines mark the trajectories of IQH states outside regime II, emanating from the charge neutrality point. The boundaries of regime II are delineated by black solid lines.
(b) Schematic phase diagram summarizing the principal transport features shown in panel (a). The purple dashed line denotes the boundary $n = 2n_{\mathrm{CDW}}$, identified from Fig.~\ref{fig2}e and Fig.~\ref{nCDW}. On the low-density side of this boundary ($n<2n_{\mathrm{CDW}}$), transport signatures indicate the coexistence of CDW order and superconductivity. On the high-density side ($n>2n_{\mathrm{CDW}}$), the coexistence of CDW order and Landau quantization gives rise to the RIQH effect. The coincidence of this phase boundary with $n = 2n_{\mathrm{CDW}}$ suggests that CDW formation reconstructs the electronic band structure into multiple sub-bands. Superconductivity is stabilized while itinerant electrons occupy the first reconstructed sub-band ($n<2n_{\mathrm{CDW}}$), whereas Landau quantization emerges once carriers begin to populate the second sub-band ($n>2n_{\mathrm{CDW}}$).
(c) Schematic illustration of the reconstructed sub-bands in the CDW state. At $n = n_{\mathrm{CDW}}$ (top), all electrons are localized within the CDW. As the carrier density increases, additional electrons remain itinerant and occupy reconstructed sub-bands. At $n = 2n_{\mathrm{CDW}}$ (bottom), the first reconstructed sub-band is completely filled by itinerant electrons. The emergence of Hall quantization only near $n = 2n_{\mathrm{CDW}}$ may indicate that Landau-level gaps are enhanced when the area per Landau level becomes commensurate with the CDW unit cell.
}
\end{figure*}

\subsection{Landau quantization and CDW order at the flat-band edge}

In conventional quantum Hall systems, the Coulomb interaction energy, $\Delta_{\mathrm{Coulomb}}$, is smaller than the Landau-level separation, $\Delta_{\mathrm{Cyclotron}}$~\cite{inti2013LL,PhysRevB.81.113301,eisenstein_stripe2001,Lilly1999stripe,Nayak_mixing}. In this regime, inter-Landau-level mixing is weak, and interaction-driven charge-density-wave order, when it occurs, is usually understood as arising from mixing of the degenerate states within a partially filled Landau level~\cite{fukuyama_yoshioka_CDW,yoshioka_lee_CDW}. Electron and hole Wigner crystal CDW states are stable when LLs are nearly empty and nearly full respectively.
Experimentally, they appear as integer quantum Hall plateaus, similar to those induced by disorder, 
because the Fermi level lies in localized states; see Fig.~\ref{LL}a.  In both cases, the Hall conductance is determined by the number of partially or fully occupied Landau levels below the Fermi energy, with each such level contributing either zero or one unit of Chern number.  The localized states at the Fermi level  within plateaus do not change the quantized Hall response, as illustrated in Fig.~\ref{LL}a.

The flat-band edge of R6G represents a distinct regime in which LL separations are not 
constant and $\Delta_{\mathrm{Coulomb}}$ exceeds the total spacing between the first $n^*$ LLs
($n^*$ is estimated below) by roughly an order of magnitude.  
In this limit, strong LL mixing~\cite{allaninfluence1984,inti2013LL,mishra2026CDW} plays a dominant role.
In the following, we present band-structure calculations near the flat-band edge of R6G, and discuss the consequences of strong Coulomb interactions in this regime and their implications for the emergence of CDW order. 

The part of the displacement-field/carrier-density phase space of R6G explored in this 
work is occupied by a spontaneously spin- and valley-polarized quarter-metal state at zero magnetic field~\cite{Li2025trashcan,Bernevig2025trashcan,Aoki2007ABC,Henck2018ABC,Shi2020ABC,Chen2020ABC,Kerelsky2021ABC,Lu2024FCI,Yoon2026quarter,Parra2025nematicQM}. Because the isospin degree of freedom is largely frozen out in the regime of interest, the following calculations are restricted to a single spin-valley flavor. 
%This suggests that the observed CDW orders primarily originate from the reconstruction and mixing of Landau orbitals, rather than from additional symmetry breaking in the spin or valley sectors.
Fig.~\ref{band} illustrates the single-particle Landau-level spectrum near the conduction-band edge of R6G.  The spectrum is obtained using the continuum model of Ref.~\cite{koshinoABC_model}, with band parameters taken from the density-functional-theory calculations of Ref.~\cite{zhangDFT_ABC_trilayer}.
Fig.~\ref{band}a shows a simplified band-structure calculation retaining only the Slonczewski-Weiss-McClure (SWMc) parameters $\gamma_0$ and $\gamma_1$, the hopping parameters for nearest-neighbor intra-layer and inter-layer hopping, and therefore neglects trigonal warping and smaller terms. Landau quantization near this flat-band edge generates a manifold of LLs that are closely spaced in energy. As shown in Fig.~\ref{band}b--c, ten or more LLs are confined within an energy window of only 2--3 meV
for fields below 10 Tesla.  By comparison, the Coulomb interaction energy is on the order of 10--30 meV in the relevant magnetic-field range. Consequently, Coulomb interactions are expected to strongly mix these nearly degenerate LLs.
%the estimate is just ~ e^2/4\pi \epsilon_0 \epsilon_{HBN} \ell_B for the range of B field 1-6 T for \epsilon_{HBN} ~ 4.5. It represents an upper bound and doesnt include and screening effects 

Our calculations reveal a level-inversion effect arising from the non-monotonic momentum dispersion at the 
band edge, such that the ordering of Landau orbitals is opposite to 
that expected for a simple parabolic band. As shown in Fig.~\ref{band}c, LLs with orbital indices $n_\mathrm{{LL}}=1$ through $7$ have lower energy than the $n_\mathrm{{LL}}=0$ LL, resulting in an inverted orbital hierarchy near the band edge.  We argue that this inverted structure plays a key role in the interaction physics.  
The number of LLs $n^*$ in the low-energy window as a function of magnetic field $B$ can be estimated by 
noting that the rapid upward dispersion of 
the conduction band begins when $\hbar v k$ is comparable to the
interlayer tunneling parameter $\gamma_1$ (i.e., for $k^*\sim \sqrt{4/3}\gamma_1/(a\gamma_0)$), where $v$ is the Dirac velocity of 
single layer graphene, and that the semi-classical area in momentum space per LL is $2\pi/\ell^2$ where $\ell$ is the magnetic length.
It follows that $n^*\sim(\ell k^*)^2 /2\sim 100/B[{\rm Tesla}]$, in agreement with Fig.~\ref{band}. 

Fig.~\ref{band}d presents a more realistic calculation that incorporates SWMc parameters $\gamma_0$, $\gamma_1$, $\gamma_2$, $\gamma_3$, $\gamma_4$ and $\delta$, thereby accounting for trigonal warping, next-nearest layer hopping and the dimer--non-dimer site-energy difference. The non-monotonic dispersion of the zero-field bands in the simplified model becomes anisotropic, resulting in three momentum-space pockets related by 120$^{\circ}$ rotations at the band edge \cite{koshinoABC_model,koshinoABCLL}.  Despite this additional complexity, Landau quantization continues to produce a manifold of nearly degenerate LLs, indicating that strong LL mixing is still expected. Trigonal warping introduces an approximate threefold degeneracy of the lowest-energy LLs, as illustrated in Fig.~\ref{band}f, corresponding to pocket-localized semiclassical orbitals
at the lowest energies that are coupled by magnetic breakdown~\cite{magnetic_breakdown1,magnetic_breakdown2}.
The level-inversion effect also persists in this more realistic model, with the $n_{\mathrm{LL}} = 0$ LL remaining elevated in energy relative to several higher-orbital LLs and the number $n^*$ of low energy LLs is 
unchanged.

These non-interacting calculations provide a useful baseline for understanding Landau quantization in the flat-band regime.  We next consider the effects of Coulomb interactions.  The unusual LL
spectrum leads to exceptionally small level spacings for the first 
$n^*$ LLs. Because the Coulomb interaction energy exceeds the separation between 
neighboring LLs, accounting for level mixing is essential.  
The correlated ground state is no longer determined by the partial filling factor 
of an isolated LL, and interaction-driven reconstructions involving multiple levels can be important. 
Indeed the observation of quantum Hall effects with 
quantization indices that differ from the filling factor by more than one, requires that strong mixing be active, as we explain below where we associate 
these reentrant integer quantum Hall states with density wave states.  The CDW
states active in R6G are qualitatively distinct from those that form within higher Landau 
levels~\cite{Fogler1996stripe,Yoshioka_shibata_higherLL}, which lead to Hall quantization 
indices adjacent to the integer part of the filling factors and 
are related to nodes in the form factors of a single LL.

The LL inversion illustrated in Fig.~\ref{band}c,f, further enhances the tendency toward CDW formation. In conventional quantum Hall systems, such as monolayer graphene, the lowest LL hosts fluid-like fractional quantum Hall states, whereas higher Landau orbitals are well known to favor CDW phases, including stripe, bubble, and Wigner-solid states~\cite{Fogler1996stripe,Koulakov1996stripe,Lilly1999stripe,Lilly1999stripe2,Du1999stripe,Pan1999stripe,Fradkin1999stripe,Eisenstein2002bubble,Fogler2002stripe,Goerbig2003RIQH,Goerbig2004electronsolid,Xia2004bubble,Deng2012reentrant,Gervais_3rdLL,Liu2012reentrant,Halperin2020FQHE,Chen2019RIQH,Friess2018bubblestripe}. Near the flat-band edge of R6G, however, the $n_{\mathrm{LL}} = 0$ LL is inverted above several higher-orbital LLs within a narrow energy window. Occupying the $n_{\mathrm{LL}} = 0$ LL therefore carries a single-particle energy penalty relative to the nearby higher-orbital LLs. As a result, Coulomb-driven electronic states may preferentially draw weight from nearby higher Landau orbitals, making CDW order particularly competitive.

In Fig.~\ref{LL}b, we provide a schematic illustration of CDW formation through the mixing and reconstruction of eight LLs. This schematic demonstrates how strong LL mixing can generate
states with different strength of CDW order that are distinguished by their total Chern numbers. 
As a representative example, we consider the case in which eight nearly degenerate LLs are present in the low-energy region, and the LL filling factor $\nu_{\mathrm{LL}} = 4$, so that four of the eight LLs are occupied. In the absence of CDW order each LL carries a Chern number of one. Consequently, occupation of four LLs gives rise to a Hall conductivity quantized at $\sigma_{xy}=4e^2/h$.
Generally speaking CDW order will lead to a periodic pattern of charge density peaks.
When these peaks are formed by a single electron, Coulomb self-interaction is absent \footnote{In
Hartree-Fock mean-field theory this absence of self-interaction is manifested by a nearly perfect
cancellation between Hartree and Fock mean fields.}.  When more than one LL is available in 
a narrow region of kinetic energy, the maximum local charge density is increased and the 
charge-density peaks can become narrower, lowering the total energy.

When the LLs are reconstructed by Coulomb interactions to form well-separated charge-density 
peaks, CDW order can redistribute Chern numbers among the participating levels. 
A similar Chern-number redistribution is familiar from non-interacting reconstructed band structures, in
which band inversions and gap closings can transfer Chern number between bands~\cite{wu_topological_TMD,morales_duran_Adiabatic,mishra2026CDW}. 
Related Chern-number transfer also occurs in Hofstadter systems, in which
a periodic potential reconstructs LLs into magnetic sub-bands with redistributed Chern numbers~\cite{hfstadter_paper,TKNN_1982,Allan_1984_LLperpot}. Here, by contrast, the reconstruction is driven by interaction-induced CDW order, and owing to the negligible LL energy 
separations can occur between LLs and not only
between subbands of a particular LL.  In the schematic picture we focus on 
integer total filling factors, for which we expect the largest gaps and the most stable CDW states.
When a fully occupied LL is reorganized into a periodic distribution of localized charges, its Chern number is reduced to zero (shown as hashed) and transferred to another level in the reconstructed manifold (shown as orange). Consequently, the observed Hall conductivity is determined not simply by the total number of occupied LLs, but by the subset of occupied LLs that remain topologically nontrivial and do not participate in CDW formation. In this sense, the difference between the Hall quantization expected from the non-interacting LL filling and the experimentally observed Hall quantization provides a direct measure of the extent to which the electronic ground state has been reconstructed by CDW order.

As illustrated in Fig.~\ref{LL}b, the formation of CDW order progressively reduces the number of occupied LLs with non-zero Chern number. Consequently, as more carriers become incorporated into the CDW state, the observed Hall quantization is correspondingly reduced.  This is our 
interpretation of the observation in Fig.~\ref{fig2}a--b which 
exhibits a Hall conductivity quantized at $4e^2/h$ despite occurring at an effective LL filling of approximately eight. Within the framework of Fig.~\ref{LL}b, this observation is consistent with a CDW state that incorporates four occupied LLs, while the remaining four occupied LLs retain nonzero Chern number and contribute to the quantized Hall response.
This mechanism provides a natural framework for understanding the sequence of RIQH states observed in Fig.~\ref{fig1}e--f, each of which exhibits a distinct Hall quantization. In the extreme limit, illustrated in Fig.~\ref{LL}b-(iii), all occupied LLs become incorporated into the CDW state, yielding a fully charge-ordered phase with vanishing Hall conductivity.

A slightly different picture emerges when $1< n/n_{CDW} < 2$.   According to Fig.~\ref{fig2}e and Fig.~\ref{nCDW}, the density of localized electrons varies continuously with $n$. 
As a result, localized and itinerant electrons coexist throughout this regime. However, unlike the situation illustrated in Fig.~\ref{LL}b, a perpendicular magnetic field does not open a well-defined energy gap between LLs of the itinerant electrons. Consequently, transport remains non-quantized despite the presence of the underlying CDW order.

One possible explanation is that the CDW forms a honeycomb lattice that reconstructs the electronic structure into two flat bands. For $1 < n/n_{\mathrm{CDW}} < 2$, the itinerant electrons occupy the lower-energy band, which is less dispersive than its higher-energy counterpart. 
The resulting LLs are therefore expected to have a smaller energy separation, preventing the opening of a robust mobility gap. Consequently, Hall quantization is absent even though the underlying CDW order persists.

\begin{figure*}
\includegraphics[width=0.89\linewidth]{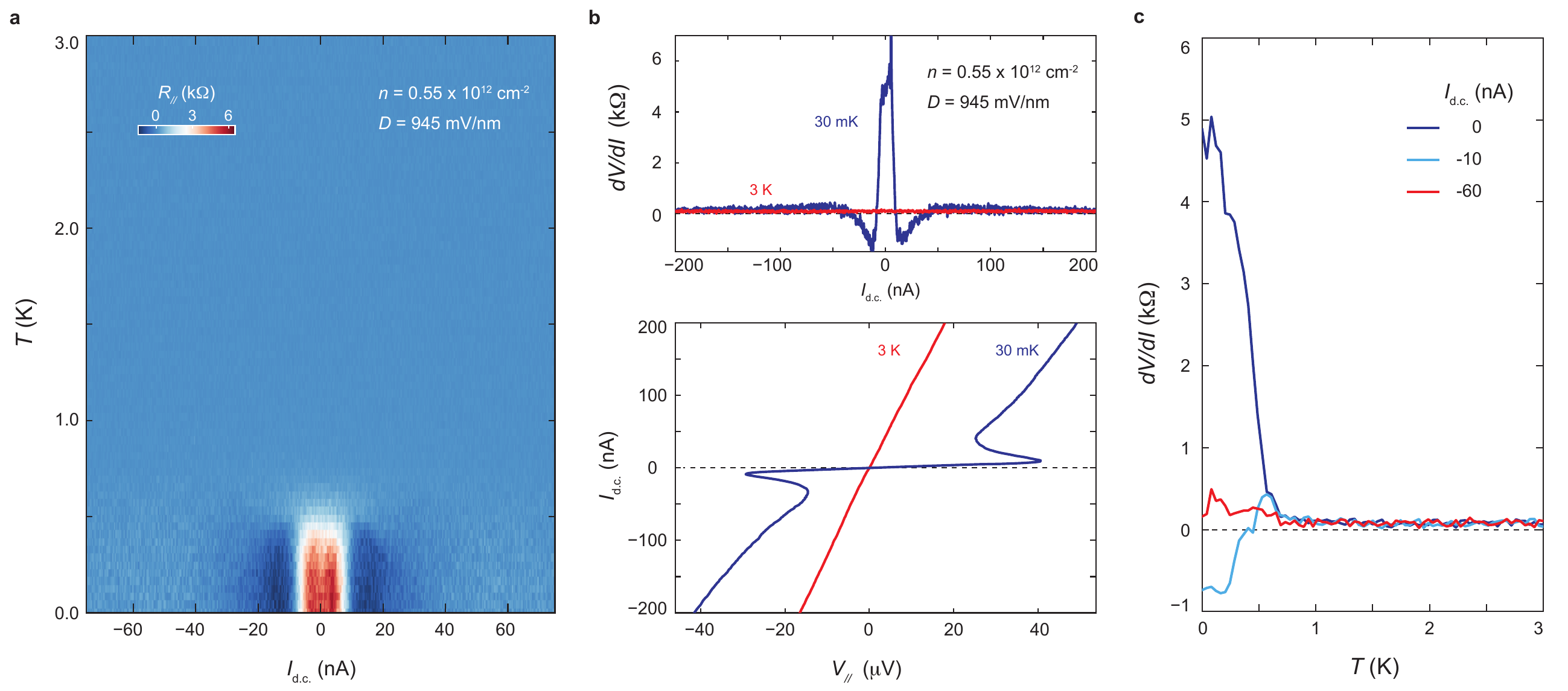}
\caption{\label{IVT} \textbf{Thermal melting and current-driven breakdown of the charge crystal.} All measurements are performed at the same $n,D$ values as in Fig.~\ref{fig2}f.
(a) Differential resistance $dV/dI$ as a function of $I_{\mathrm{d.c.}}$ and temperature $T$. 
(b) $dV/dI$ as a function of $I_{\mathrm{d.c.}}$ (top), and the corresponding integrated $I$–$V$ curves (bottom), highlighting nonlinear transport at $30$~mK (blue) compared with the linear response at $3$~K (red).
(c) Temperature line cuts extracted from the map in panel (a) at $I_{\mathrm{d.c.}} = 0$ (dark blue), $-10$~nA (light blue), and $-60$~nA (red).
}
\end{figure*}

\begin{figure*}
\includegraphics[width=0.99\linewidth]{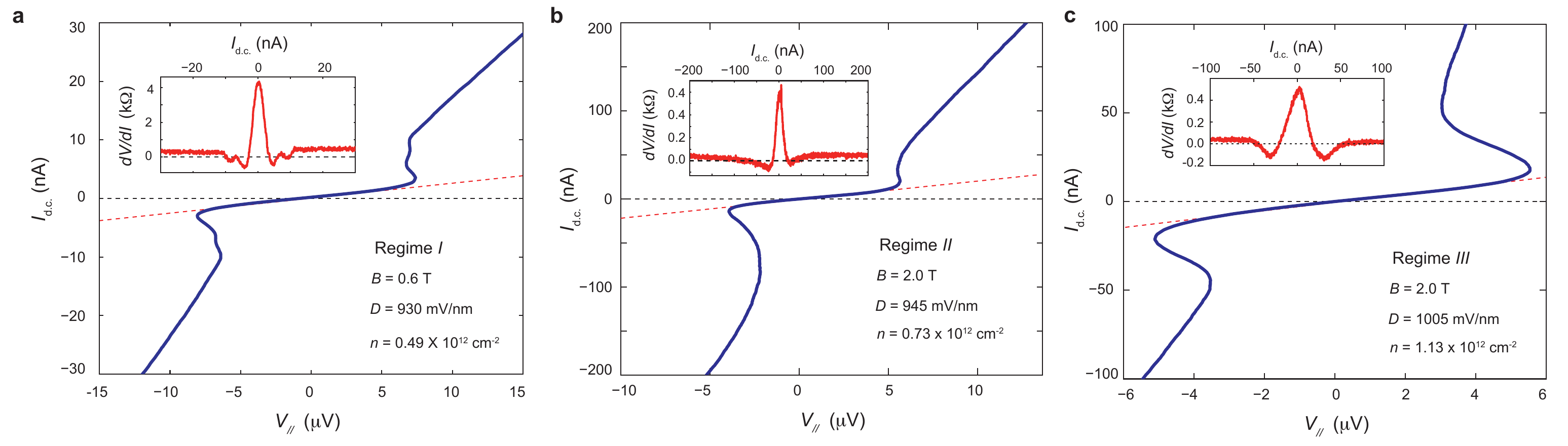}
\caption{\label{CDW} \textbf{Current-driven breakdown of CDW orders in regimes I--III.} 
$I$--$V$ curves measured in (a) regime I, (b) II, and (c) III. Insets show $dV/dI$ as a function of $I_{\mathrm{d.c.}}$. Each $I$--$V$ curve exhibits hallmark signatures of a Wigner-solid-like phase, where a highly resistive response at low bias transitions into a conductive response at large $I_{\mathrm{d.c.}}$. The transition is marked by pronounced negative dips in the differential resistance, which corresponds to the current-driven breakdown of the crystalline order. Moreover, the slope of the $I$--$V$ curve in the low-bias response, marked by the red dashed line, indicates that the state is not fully insulating, consistent with a population of residual itinerant carriers coexisting with the charge crystal.
}
\end{figure*}

\begin{figure*}
\includegraphics[width=0.99\linewidth]{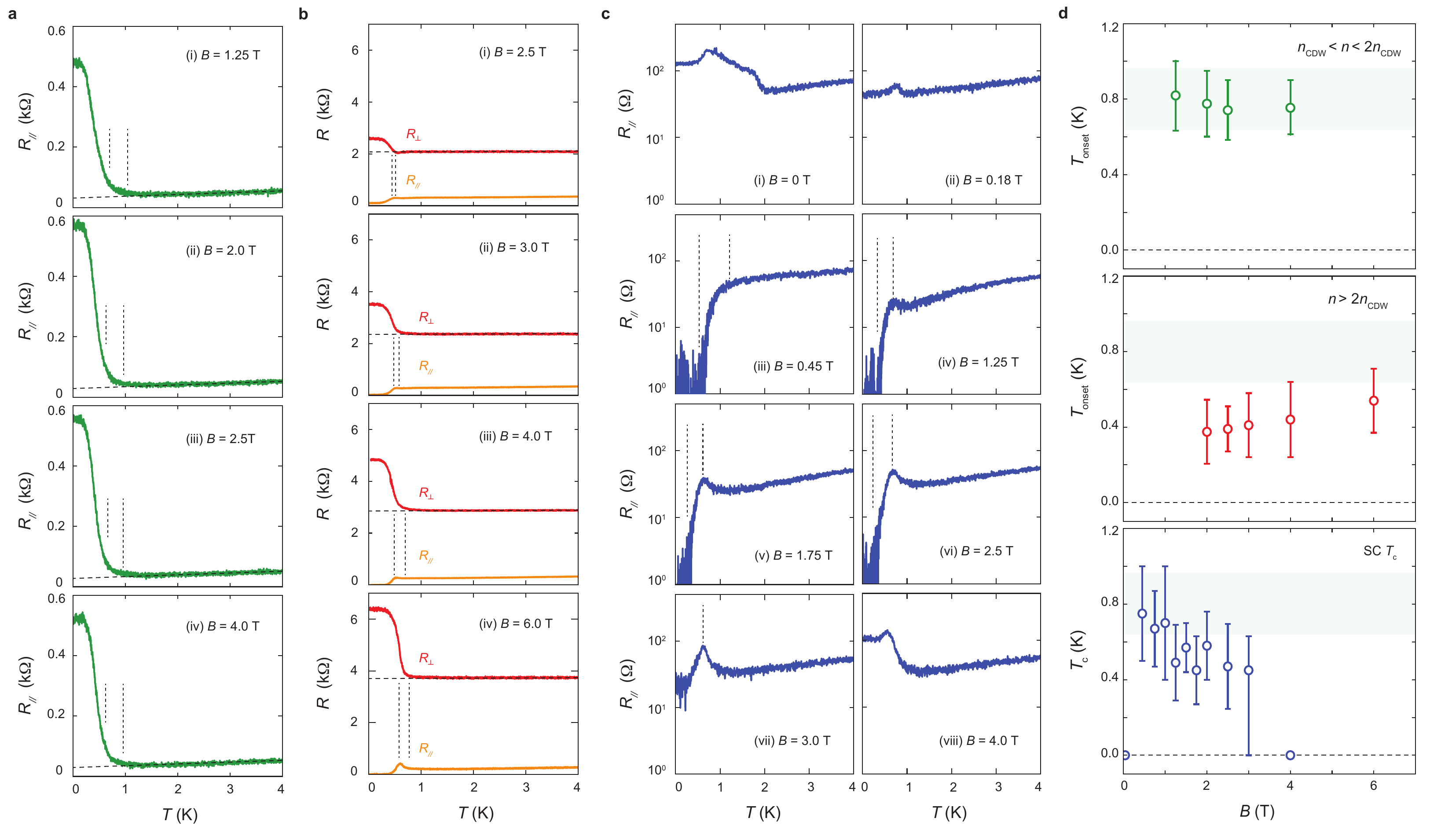}
\caption{\label{onset} 
\textbf{Temperature-driven onset of CDW orders and superconductivity.} 
All three phases observed within regime II exhibit a sharp onset upon cooling. Panels (a--c) show the temperature dependence of the transport response at different \Bperp\ for (a) the CDW order at $1< n/n_{CDW} < 2$, (b) the CDW order at $n/n_{CDW} > 2$,  and (c) the superconducting phase. In each panel, onset temperatures are indicated by vertical dashed lines. 
In (a), the onset of the insulating state is defined based on the RIQH response. The vertical dashed lines mark the temperatures at which \Rperp\ departs from the high-temperature Hall resistance and \Rpara\ develops a pronounced peak. 
In (b), the onset of the RIQH effect is defined as the temperature at which \Rpara\ deviates from its high-temperature trend; the two black dashed lines indicate the uncertainty range used to determine the error bar. 
In (c), the two dashed lines mark the temperatures at which \Rpara\ first exhibits a downturn and subsequently falls to zero. 
(d) Onset temperatures of all three phases plotted as a function of $B_{\perp}$.
}
\end{figure*}

\begin{figure*}
\includegraphics[width=0.99\linewidth]{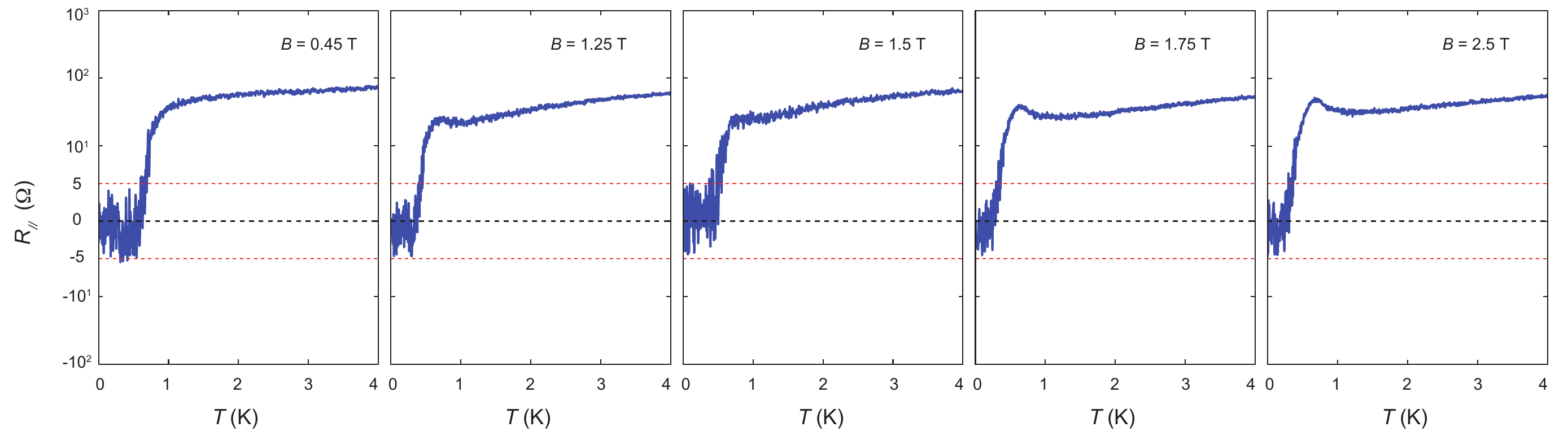}
\caption{\label{RT_SC} \textbf{\Rpara--$T$ traces of the superconducting phase at different \Bperp.} 
Temperature dependence of the transport response of the superconducting phase in regime II measured at different \Bperp, plotted on a bi-symmetric logarithmic scale. In each panel, red horizontal dashed lines mark $\pm 5\,\Omega$. The measurement noise floor is approximately $\pm 3\,\Omega$. For each value of \Bperp, \Rpara\ measured at low temperature remains within the noise band around zero. Together with the vanishing transverse resistance (see Fig.~\ref{fig3}b--d), this response indicates a diverging conductivity across the sample, consistent with the expected transport signature of a superconducting phase.
}
\end{figure*}

\subsection{Re-entrant quantum Hall effect at $n/n_{CDW} > 2$}

The RIQH states display transport signatures resembling those observed in semiconductor quantum wells, such as re-entrant integer plateaus in the Hall conductivity, yet an important distinction remains.

In semiconductor quantum wells and monolayer graphene~\cite{Eisenstein2002bubble,Fogler2002stripe,Goerbig2003RIQH,Goerbig2004electronsolid,Xia2004bubble,Deng2012reentrant,Gervais_3rdLL,Liu2012reentrant,Halperin2020FQHE,Chen2019RIQH}, RIQH states arise from the formation of bubble phases, in which charge carriers in a partially filled LL localize into a crystalline structure (see Fig.~\ref{LL}a) ~\cite{Fogler1996stripe}. These bubble phases occur at specific fractional LL fillings and exhibit Hall quantization identical to that of the nearest integer quantum Hall (IQH) plateau. In this conventional hierarchy, LL formation establishes the fundamental platform, while CDW order emerges as secondary instabilities.

In R6G, this hierarchy is inverted due to the strong LL mixing induced at the flat-band edge. As a result, LL formation itself becomes reconstructed by the emergence of CDW order, leading to several striking consequences.  

First, the number of charge carriers participating in the CDW order—defined by the density separation between a re-entrant Hall plateau and the IQH state with the same Hall quantization—corresponds to the combined degeneracy of multiple LLs. This provides strong evidence that the CDW order originates from the strong mixing of several LLs, rather than from a partially filled isolated LL.  

Second, despite emerging from the flat-band edge, the RIQH states appear at lower magnetic fields than the surrounding IQH states, as shown in Fig.~\ref{nB_SC}c. This observation indicates that the energy scale associated with the CDW order, driven predominantly by Coulomb interactions, exceeds the cyclotron gap of the surrounding phase space, where the underlying electronic band remains more dispersive.

While the formation of CDW order reduces the number of charge carriers contributing to transport, the total charge participating in the CDW order, and hence its lattice constant, is not fixed.  

The RIQH states form well-defined trajectories in the $n$--\Bperp\ plane. As shown in Fig.~\ref{mismatch}, the slopes of these trajectories exhibit pronounced deviations from the expectation based on the quantization of the re-entrant Hall plateaus~\cite{Streda1983streda,Huang2025streda}. Such deviations indicate that, upon field-effect doping, charge is continuously redistributed between the CDW order and the underlying LLs. Consequently, the total charge associated with the CDW order, and therefore its lattice constant, evolves continuously across the $n$--\Bperp\ plane.

\subsection{CDW order at $1< n/n_{CDW} < 2$}

The CDW order is observed across multiple regimes of the low-temperature phase space in R6G. It is identified through a characteristic current--voltage response exhibiting negative differential resistance, indicative of current-driven breakdown of the crystalline state. Similar nonlinear $I$--$V$ characteristics were previously observed in Wigner solids formed in semiconductor quantum wells~\cite{Csathy2007CDW}. In R6G, however, the $I$--$V$ curves retain a finite slope near zero bias. We attribute this finite low-bias conductance to residual carriers that remain itinerant in the presence of the CDW order.

At zero bias, the \Rpara--$T$ traces in Fig.~\ref{fig2}h provide evidence for a thermal melting transition of the CDW order. The nonlinear $I$--$V$ characteristics evolve consistently across this transition. Fig.~\ref{IVT}a plots the differential resistance $dV/dI$ as a function of $I_{\mathrm{d.c.}}$ and temperature $T$, revealing a temperature-driven transition near $T=0.5$~K. Above this transition, the resistive response near zero bias becomes highly conductive, consistent with melting of the underlying CDW order. Simultaneously, the nonlinear $I$--$V$ response associated with current-driven breakdown disappears, and the transport becomes fully ohmic, indicative of a conventional metallic state (Fig.~\ref{IVT}b).

The negative dip in $dV/dI$ is interpreted as a signature of current-driven breakdown of the CDW order~\cite{Csathy2007CDW}. As illustrated in Fig.~\ref{IVT}c, the negative dip in $dV/dI$ at $I_{\mathrm{d.c.}}=-10$~nA emerges simultaneously at the thermal transition, establishing a direct link between the nonlinear $I$--$V$ response and the CDW order. Moreover, Fig.~\ref{IVT}c shows that $dV/dI$ becomes largely temperature independent at large d.c. bias beyond the breakdown threshold. This behavior indicates that, once the breakdown occurs, the CDW order is destroyed and the sample transitions into a conductive metallic state.

This distinctive nonlinear transport response, marked by negative dips in $dV/dI$, is consistently observed throughout regimes I--III (Fig.~\ref{CDW}). By contrast, as shown in Fig.~\ref{linearIV}, the $I$--$V$ curves become fully linear once the density is tuned outside these regimes. We therefore identify regimes I--III as the low-temperature phase space in which stable CDW order forms.

Figures~\ref{R_n_temp} and \ref{PnBT} plot the transport response around regime II at $1.5$~K. Above the melting transition, regime II becomes indistinguishable from the surrounding metallic phase. This observation further supports the interpretation that regimes I--III are defined by the emergence of the underlying CDW order.

In addition to the nonlinear $I$--$V$ response, the emergence of the CDW order in regimes I--III is also reflected in the evolution of transport anisotropy. As shown in Fig.~\ref{nD_anisotropy}a, the phase space surrounding regimes I--III exhibits pronounced anisotropy at $1.5$~K. Upon cooling below $0.5$~K, where the CDW order develops, the anisotropy evolves qualitatively differently. As shown in Fig.~\ref{nD_anisotropy}b, although the regions between regimes I--III remain anisotropic at $B=0$, the transport response within regimes I--III becomes largely isotropic. This observation indicates that the CDW order in these regimes is predominantly isotropic, distinguishing it from stripe order.

\subsection{The onset temperature of CDW order and superconductivity}

Within regime II, transport measurement reveals three distinct behaviors: a resistive state at $1 < n/n_{CDW} < 2$, re-entrant quantum Hall effect at $n/n_{CDW} > 2$, and superconductivity.  These phases share a common origin rooted in an underlying CDW order and its coexisting itinerant charges. Here, we explore the temperature-driven onset of these phases and examine their evolution with varying \Bperp. 

Fig.~\ref{onset}a--c plots $R$--$T$ traces measured from the resistive, RIQH, and superconducting phases within regime II at different \Bperp\ values. 

For the insulating state, the locations of the vertical dashed lines are determined using the criteria shown in Fig.~\ref{Insulating_onset}. For the RIQH state, the vertical dashed lines denote the temperatures at which \Rpara\ and \Rperp\ deviate from their high-temperature behavior, respectively. For the superconducting transition, the higher-temperature dashed line marks the downward turn in the $R$--$T$ trace, signaling the onset of superconducting fluctuations, whereas the lower-temperature dashed line marks the temperature at which \Rpara\ vanishes to zero.

The combination of these two vertical dashed lines mark the temperature range of the onset transition, which define the error bars in Fig.~\ref{onset}d.  

Fig.~\ref{onset}d plots the onset temperatures extracted from Fig.~\ref{onset}a--c. The onset of insulating behavior is marked by green circles in the top panel, which is found to be largely independent of \Bperp. This temperature denotes the formation of the CDW order. The horizontal green stripe serves as a guide to the eye, highlighting this characteristic temperature scale.

As illustrated in the middle panel of Fig.~\ref{onset}d, the onset temperature of the RIQH state increases slightly with increasing \Bperp, consistent with the expected behavior of quantum Hall states. However, unlike conventional quantum Hall states, which exhibit a gradual temperature dependence (see Fig.~\ref{RT_QHE}), the RIQH states display sharply defined temperature transitions, which corresponds to the melting transition of the underlying CDW order.

The superconducting phase within regime II exhibits a non-monotonic dependence on \Bperp. At zero field, the entire range of regime II remains resistive down to the base temperature of the dilution refrigerator (panels i and ii in Fig.~\ref{onset}c). However, as \Bperp\ increases, a superconducting transition emerges within this regime. Beyond \Bperp $= 3$~T, the superconducting state progressively weakens. At $4$~T, the $R$--$T$ traces recover an insulating response, reflecting the temperature dependence of the underlying CDW order.

\begin{figure*}
\includegraphics[width=0.99\linewidth]{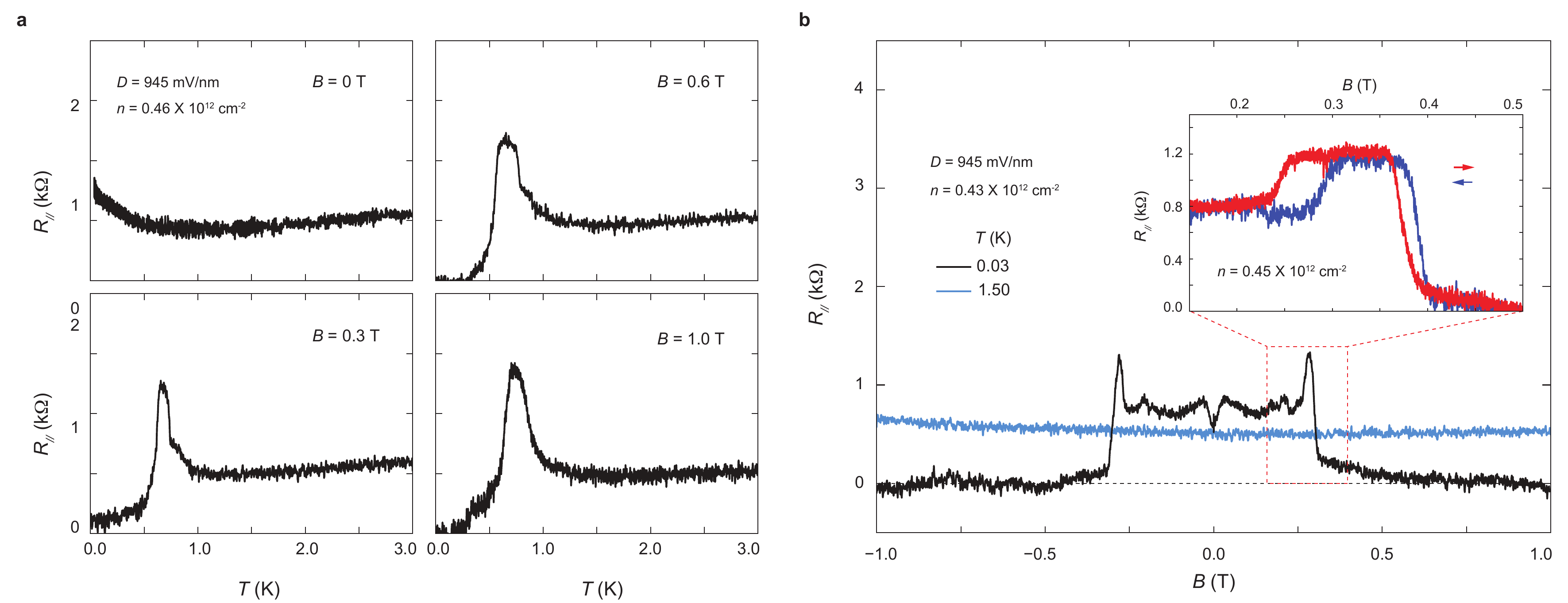}
\caption{\label{B_dependence} \textbf{\Bperp\ dependence of superconductivity.} 
(a) Temperature dependence of \Rpara\ measured at different values of $B_{\perp}$ within regime I, at $n = 0.46 \times 10^{12}$ cm$^{-2}$ and $D = 945$ mV/nm. The superconducting state is stabilized by a perpendicular magnetic field.
(b) $R_{\parallel}$ as a function of $B_{\perp}$ measured at the same location. A sharp field-driven transition separates the superconducting and resistive states. Inset: enlarged view near the transition, showing pronounced hysteresis between forward and reverse field sweeps. 
}
\end{figure*}

As shown in the bottom panel of Fig.~\ref{onset}d, the superconducting transition temperature displays an overall downward trend with increasing \Bperp. This trend is consistent with the expected behavior of a two-dimensional superconductor in a perpendicular magnetic field. The persistence of a robust superconducting phase beyond \Bperp $= 3$~T suggests that the order-parameter structure may be unconventional in nature.

The superconducting phase is absent within regime II at zero field but becomes stabilized by a finite perpendicular field. This behavior likely reflects the \Bperp-dependence of the underlying CDW order, providing further evidence that the stability of the superconducting phase is bounded by the stability window of the crystalline order.

It is also worth noting that superconducting and RIQH states coexist over an extended range of \Bperp\ (see also Fig.~\ref{nB_SC}). Both phases emerge within regime II and occupy adjacent regions in the $n$--$D$ parameter space. This coexistence is highly unusual and reflects an unprecedented phenomenon in which superconductivity arises from LL formation.

In addition, we note that the zero-field $R$--$T$ trace in Fig.~\ref{onset}c is distinct from the insulating response shown in Fig.~\ref{onset}a. Instead of exhibiting a sharp onset below a well-defined critical temperature, the zero-field $R$--$T$ trace shows only a modest step in \Rpara\ around $2$~K. 
These observations are consistent with a scenario in which the CDW order is stabilized by a perpendicular magnetic field, which in turn appears to provide the necessary condition for stabilizing the superconducting phase.

\subsection{Coexisting CDW and superconductivity in regime I}

In regime I, CDW and superconducting phases are stabilized by a non-zero \Bperp. Fig.~\ref{linecut}a shows the $n$--$D$ map of \Rpara\ measured at \Bperp $=0.6$~T, centered around resistive regime I. In the chosen color scale, high and low resistance are represented by red and dark blue, respectively. This reveals a superconducting phase along the boundary of the resistive state.

Both the superconducting and resistive states exhibit well-defined onset behavior with varying temperature. 
Fig.~\ref{fig4}d plots the temperature dependence of \Rpara\ measured in the superconducting and resistive states, at the locations marked by the orange and blue circles in Fig.~\ref{linecut}, respectively. Above $T \sim 1$~K, the two traces are nearly indistinguishable. Upon cooling, however, a clear bifurcation emerges around $T \sim 0.7$~K. In particular, \Rpara\ exhibits a pronounced peak immediately above the superconducting transition, matching the onset observed in the resistive state. Such temperature dependence closely resembles the behavior observed in Fig.~\ref{fig3}e, implying a similar interplay between superconductivity and CDW order. In this context, the emergence of the CDW order coincides with the resistive onset around 0.7 K.

The presence of a CDW order in regime I is further supported by its melting transition, which gives rise to pronounced thermal hysteresis between cooling and warming measurements (Fig.~\ref{fig4}e). Similar hysteresis is also observed above the superconducting transition, as shown in Fig.~\ref{fig4}f. Notably, the onset of hysteresis coincides with the emergence of the resistive peak. 

These observations establish a temperature hierarchy: upon cooling, CDW order forms first, followed by the onset of superconductivity at lower temperature. The hysteresis observed in the superconducting transition therefore reflects the superheating and supercooling of the underlying CDW order with which it coexists. This hierarchy is consistent with our observations in regime II, as illustrated in Fig.~\ref{fig3}.

Apart from the coexistence with CDW order, the superconducting phase in regime I displays an intriguing dependence on \Bperp. This is shown in Fig.~\ref{B_dependence}a, which plots $R$--$T$ traces at different \Bperp\ from a fixed $n$ and $D$ value. 
At finite \Bperp, the emergence of superconductivity is consistently observed immediately below a sharp resistive peak, signaling the formation of CDW order. By contrast, at \Bperp\ $=0$ T, superconductivity is absent, and the $R$--$T$ trace exhibits only a weak upturn at low temperature. 

We note that a similar dependence on \Bperp\ is observed for the superconducting phase in regime II (see Fig.~\ref{onset}). This \Bperp-dependence stands in sharp contrast to the conventional antagonism between superconductivity and magnetic fields. The common feature of these superconducting phases is their coexistence with CDW order, suggesting a natural interpretation: the stability of superconductivity is intimately tied to that of the CDW order, which itself is stabilized only in the presence of a perpendicular magnetic field.

Fig.~\ref{B_dependence}b further probes the role of \Bperp\ by tracking the evolution of \Rpara\ as a function of \Bperp, measured at a fixed $(n,D)$ point that lies within the superconducting phase at \Bperp $=0.6$~T. A field-driven transition is observed near $0.3$~T, where a sharp resistance peak separates the superconducting state at higher \Bperp\ from a metallic state at lower \Bperp. The form of this peak closely resembles that of the temperature-driven superconducting transition, in which a resistance peak separates the low-temperature superconducting phase from the high-temperature metallic state.  

Strikingly, the field-driven transition exhibits phenomenology closely analogous to the temperature-driven transition. As shown in the inset of Fig.~\ref{B_dependence}b, the resistance peak displays pronounced hysteresis between slow forward and reverse sweeps of \Bperp. This observation indicates that the field-driven transition is first order in nature, consistent with a quantum melting transition of the CDW order. Within this picture, the CDW order is progressively suppressed with decreasing \Bperp, naturally accounting for the unconventional magnetic-field dependence of the superconducting phase.

\section*{acknowledgments}

J.I.A.L. wishes to acknowledge helpful discussions with Andrea Young, Yao Wang, and Fan Zhang. This material is based on the work supported by the Air Force Office of Scientific Research under award no. FA9550-23-1-0482.
R.Q.N., N.J.Z, and J.I.A.L. acknowledge support from the Air Force Office of Scientific Research. E.M. acknowledge support from U.S. National Science Foundation under Award DMR-2143384.  K.W. and T.T. acknowledge support from the JSPS KAKENHI (Grant Numbers 21H05233 and 23H02052) and World Premier International Research Center Initiative (WPI), MEXT, Japan. 
Part of this work was enabled by the use of pyscan (github.com/sandialabs/pyscan), scientific measurement software made available by the Center for Integrated Nanotechnologies, an Office of Science User Facility operated for the U.S. Department of Energy. D.E.F. was supported in part by the National Science Foundation under grant No. DMR-2529089.

%\newpage
\clearpage

\newpage
\begin{widetext}
\section{Supplementary Materials}

\begin{center}
\textbf{\large Coexisting CDW and Superconducting Order in Quantizing Magnetic Fields}\\
\vspace{10pt}

Ron Q. Nguyen$^{\ast}$,
Peiyu Qin$^{\ast}$,
Hai-Tian Wu$^{\ast}$,
Sparsh Mishra,
Tobias Wolf,
Joseph Roll,
Erin Morissette,
Naiyuan J. Zhang,
Sarah Alkidim,
Kenji Watanabe, Takashi Taniguchi, Aaron W. Hui, D.E. Feldman, 

Allan MacDonald, and J.I.A. Li$^{\dag}$

\vspace{10pt}
$^{\dag}$ Corresponding author. Email: jia.li@austin.utexas.edu
\end{center}

%\noindent\textbf{This PDF file includes:}

%\newpage
\renewcommand{\vec}[1]{\boldsymbol{#1}}
\renewcommand{\thefigure}{S\arabic{figure}}
\renewcommand{\theHfigure}{S\arabic{figure}}
\renewcommand{\theequation}{S\arabic{equation}}
\renewcommand{\theHequation}{S\arabic{equation}}
\renewcommand{\thetable}{S\Roman{table}}
\renewcommand{\theHtable}{S\Roman{table}}

\setcounter{figure}{0}
\setcounter{equation}{0}
\setcounter{table}{0}

\vspace{1.1 in}

\begin{figure*}[h]
\includegraphics[width=0.95\linewidth]{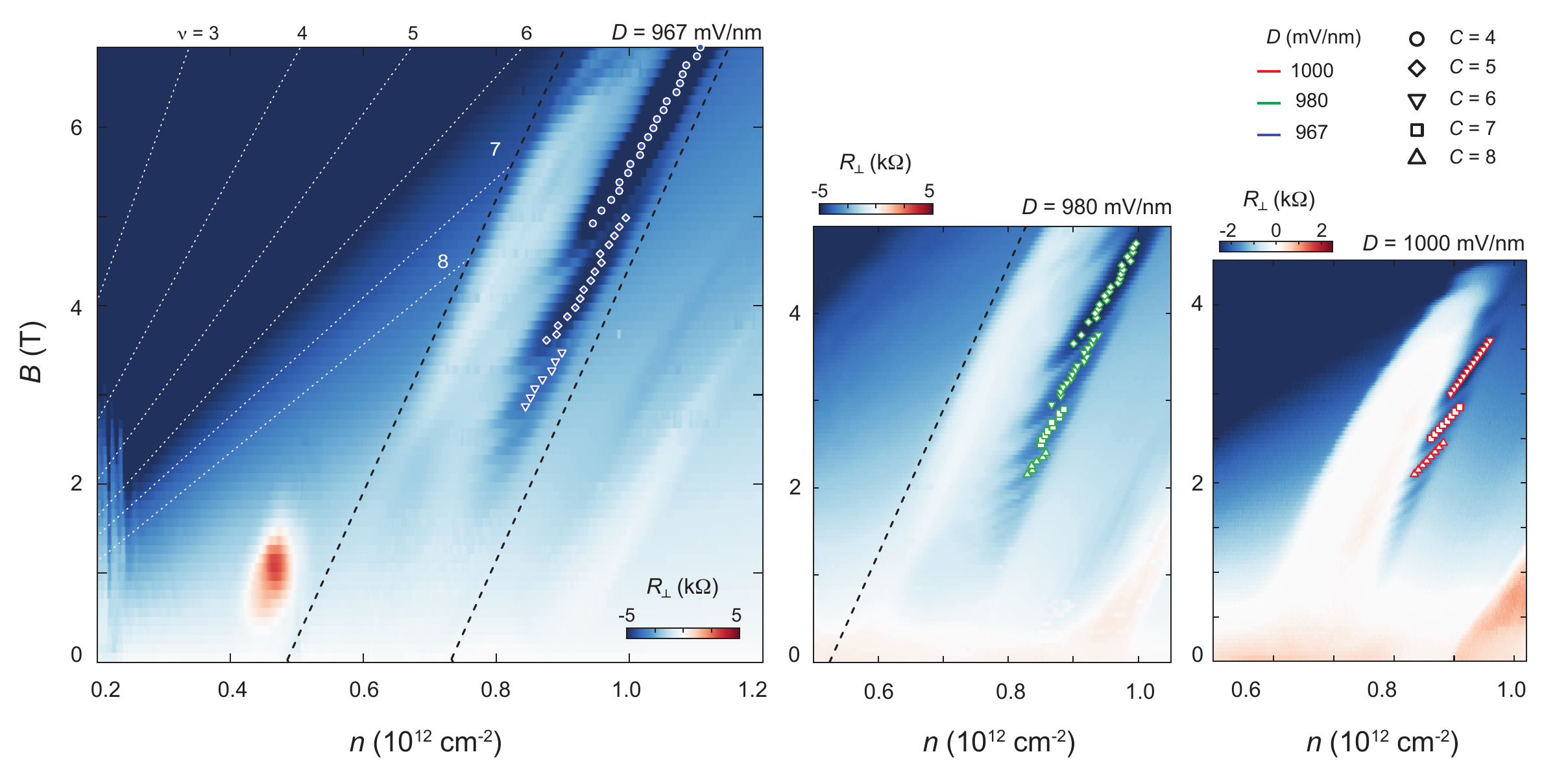}
\caption{\label{CDWanalysis}
\textbf{Unit-cell area of the CDW order.}
$n$--$B_{\perp}$ maps of \Rpara\ measured at $D = 967$~mV/nm (left), $D = 980$~mV/nm (middle), and $D = 1000$~mV/nm (right). The sequence of RIQH states appears as minima in \Rpara. Open symbols mark the $(n, B_{\perp})$ positions of the RIQH states used to extract the CDW carrier density, $n_{\mathrm{CDW}}$. The notation of the symbols is consistent with that used in Fig.~\ref{fig2}e.
}
\end{figure*}

\begin{figure*}[h]
\includegraphics[width=0.8\linewidth]{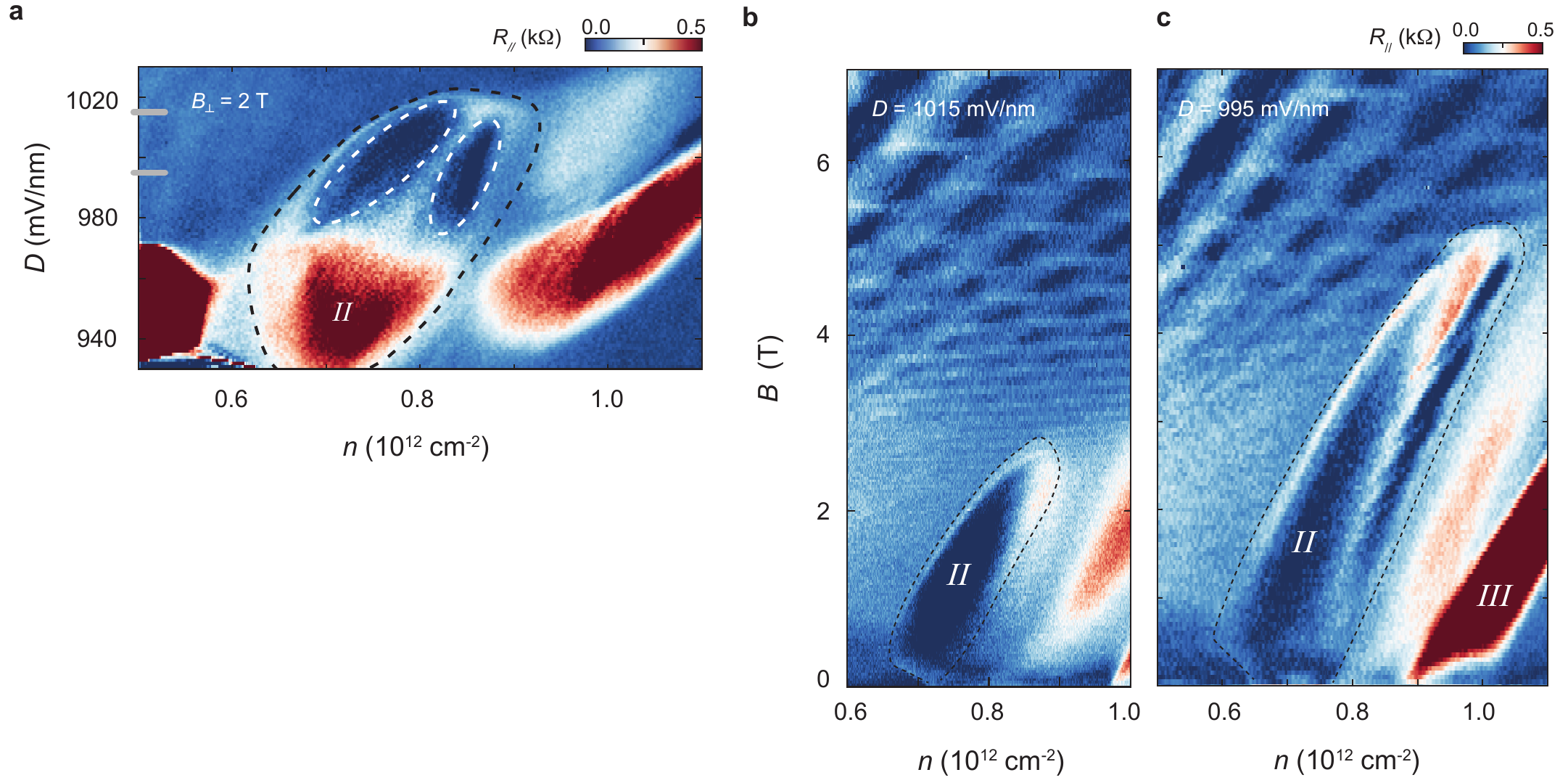}
\caption{\label{nB_SC} 
\textbf{$n-$\Bperp\ map showing superconducting phase.} 
(a) \Rpara\ plotted as a function of $n$ and $D$, measured at \Bperp\ $=2$~T. (b--c) $n$--\Bperp\ maps of longitudinal resistance  \Rpara, measured at fixed $D$ values of (b) $D = 1015\,\mathrm{mV/nm}$, and (c) $995\,\mathrm{mV/nm}$, marked by horizontal gray bars in panel (a). Each map is centered on regime II, outlined by black dashed contours.
}
\end{figure*}

\begin{figure*}[h]
\includegraphics[width=0.92\linewidth]{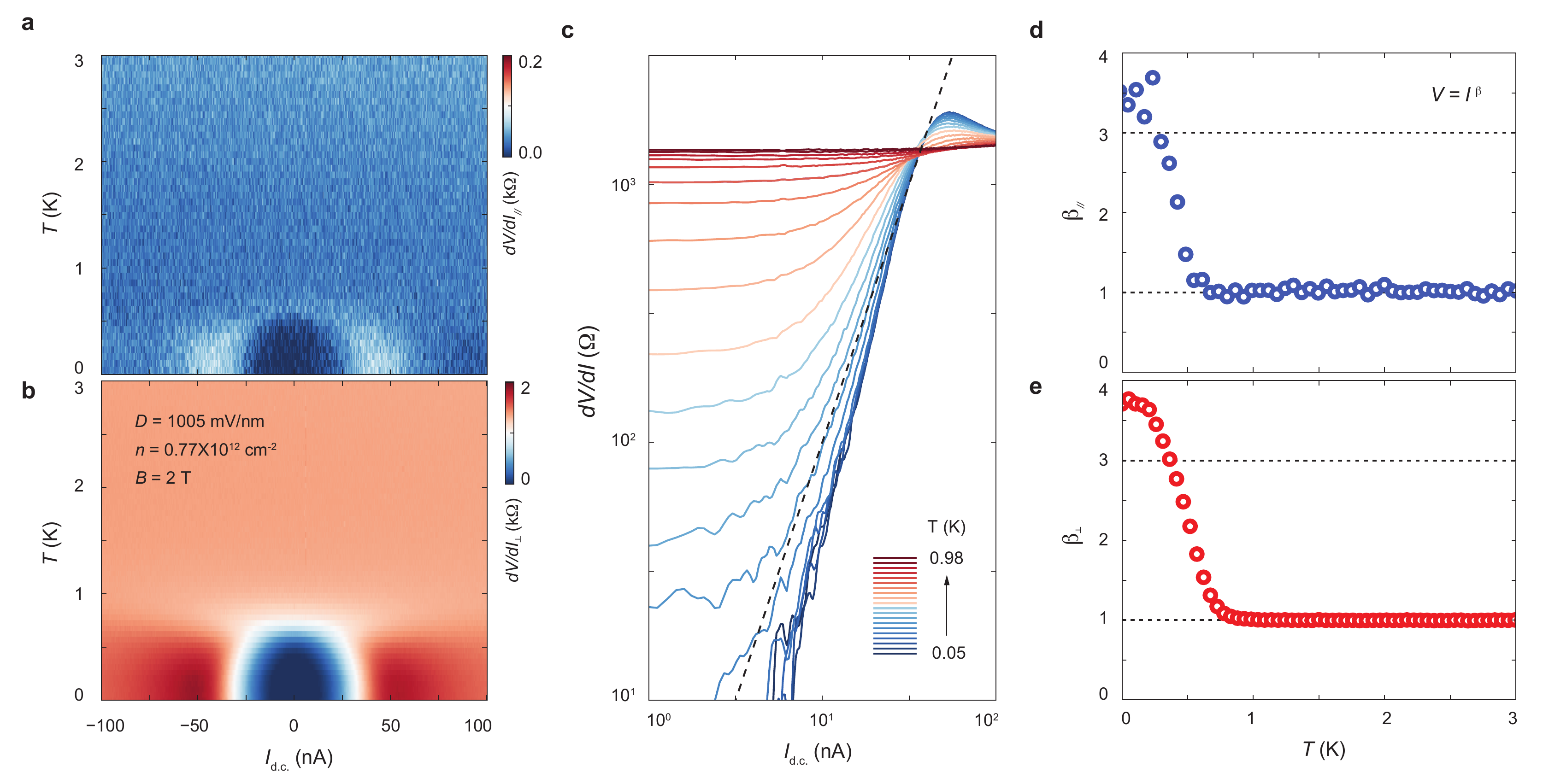}
\caption{\label{BKT} \textbf{BKT transition of the superconducting phase.} 
(a,b) Differential resistance $dV/dI$ measured in the (a) longitudinal and (b) transverse directions as a function of $I_{\mathrm{d.c.}}$ and temperature $T$. 
(c) $I$--$V$ curves measured in the superconducting phase within regime II at different temperatures. The black dashed lines indicate the power-law relation $V \propto I^{3}$. The $I$--$V$ curve exhibiting this cubic dependence defines the BKT transition temperature. 
(d,e) Power-law fits of the $I$--$V$ curves using $V \propto I^{\beta}$, from which the exponent $\beta$ is extracted for both longitudinal (d) and transverse (e) responses. The temperature dependence of $\beta$ reveals two characteristic regimes: at high temperature, $\beta = 1$, indicating fully ohmic $I$--$V$ response, while at low temperature $\beta > 3$, marking the onset of BKT-type superconducting behavior near $T \sim 0.5$~K. This temperature is consistent with the superconducting transition identified based on $R-T$ traces in Fig.~\ref{onset}. 
All measurements are performed at $B_{\perp} = 2$~T, $D = 1005$~mV/nm, and $n = 0.77\times10^{12}\,\mathrm{cm}^{-2}$.
}
\end{figure*}

\begin{figure*}
\includegraphics[width=0.8\linewidth]{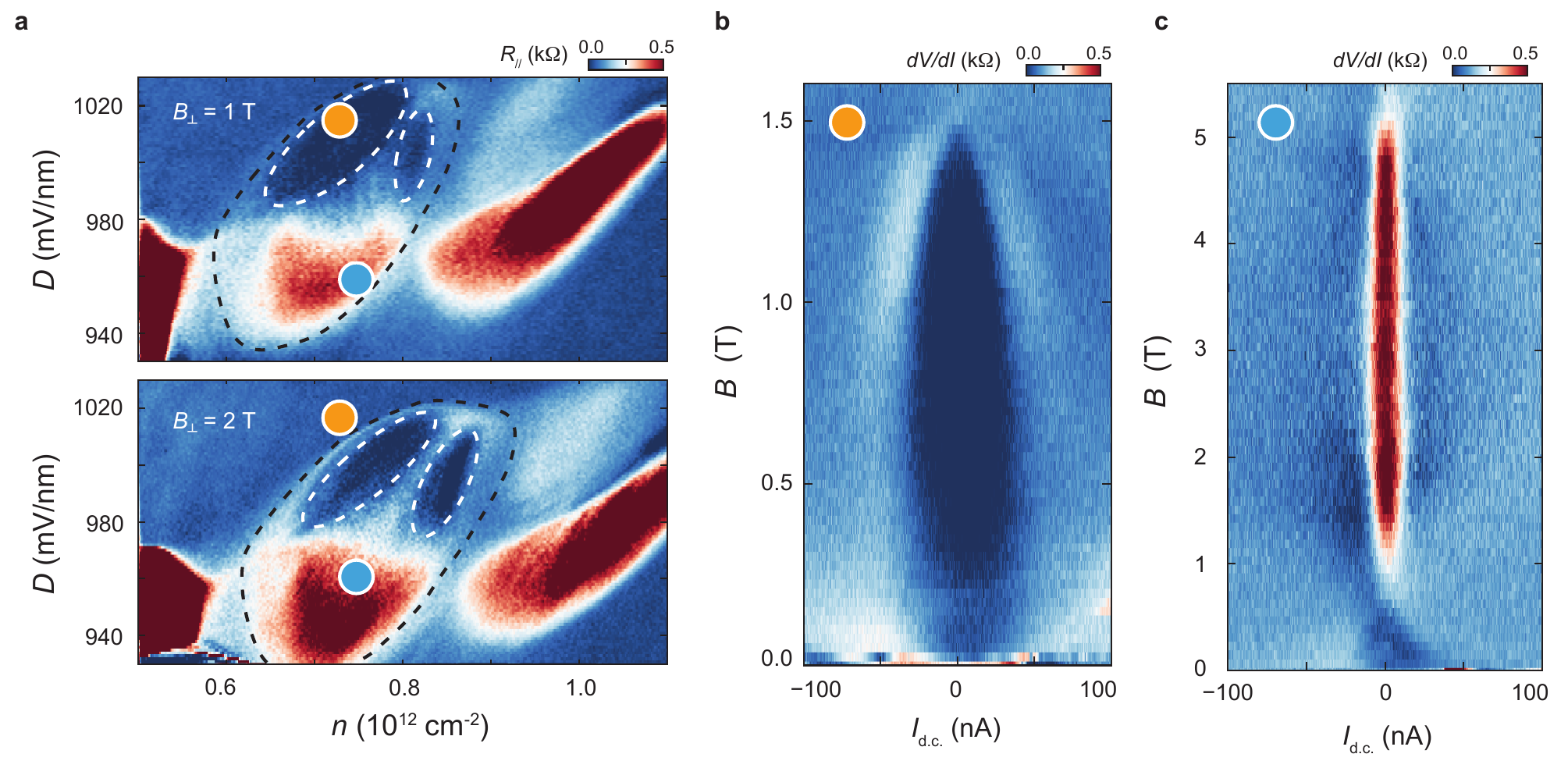}
\caption{\label{IVB} \textbf{$I$--$V$ characteristics of superconductivity and charge crystal versus \Bperp.} 
(a) Longitudinal differential resistance plotted as a function of $n$ and $D$, measured at \Bperp\ $=1$~T (top) and $2$~T (bottom). (b,c) Differential resistance $dV/dI$ plotted as a function of $I_{\mathrm{d.c.}}$ and \Bperp, measured at fixed $(n,D)$ points corresponding to (b) the superconducting state near \Bperp\ $=1$~T, marked by orange circles in panel (a), and (c) the CDW order at $1< n/n_{CDW} < 2$, marked by blue circles in panel (a). In panel (b), the apparent suppression of the superconducting response near \Bperp\ $=1.5$~T arises from the shift of the superconducting regime with increasing \Bperp. As shown in the bottom panel of (a), the orange circle lies inside the superconducting regime at \Bperp\ $=1$~T, but outside the superconducting regime at \Bperp\ $=2$~T. This shift is also apparent in Fig.~\ref{fig3}a. In panel (c), the $I$--$V$ characteristics are in excellent agreement with those shown in Fig.~\ref{fig2}f, exhibiting a resistive response at zero bias that transitions to a highly conductive response at $I_{d.c.} > 10$~nA.  This transition is accompanied by pronounced negative $dV/dI$, 
a hallmark of current-driven breakdown of the CDW order.
}
\end{figure*}

\begin{figure*}
\includegraphics[width=0.55\linewidth]{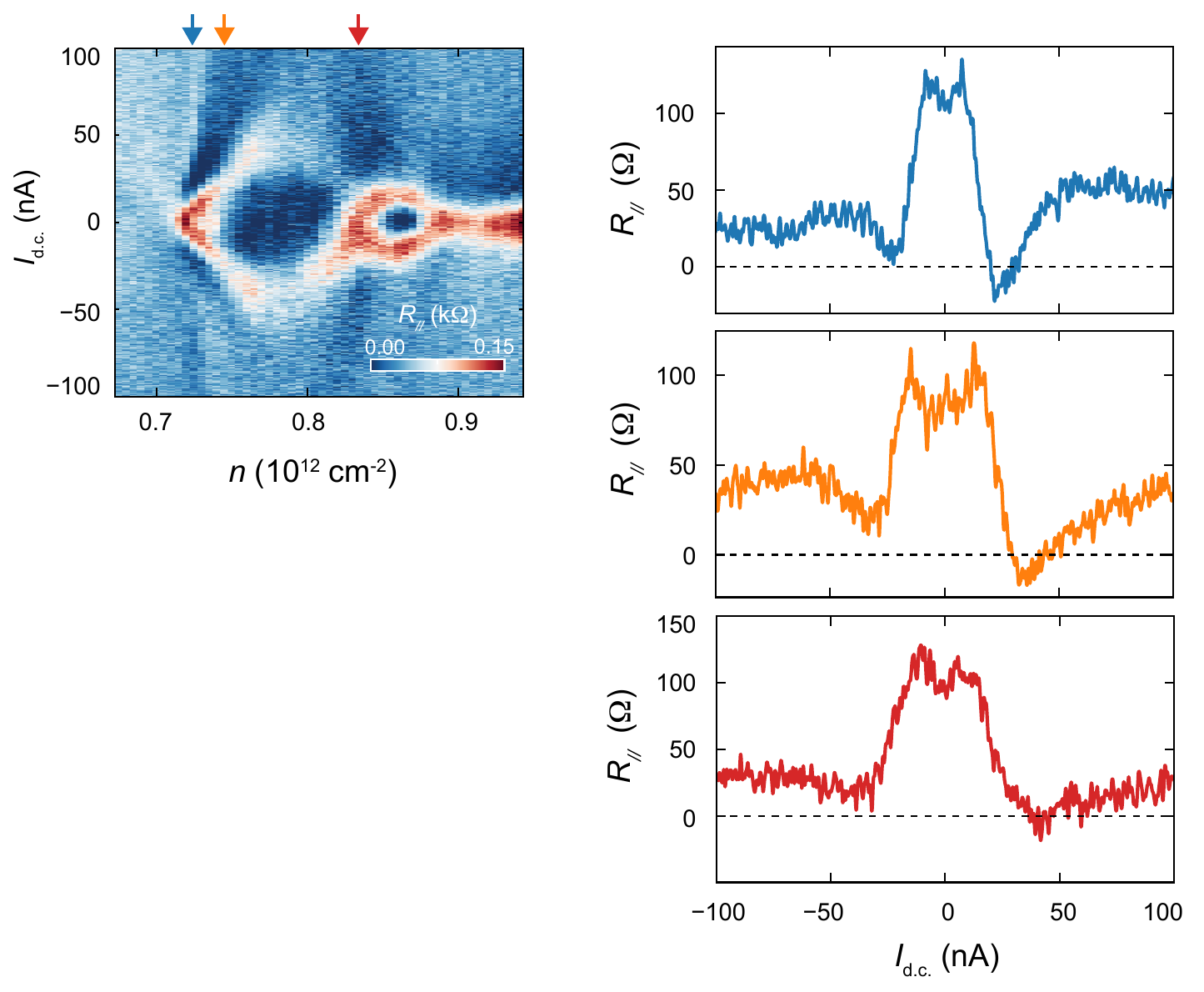}
\caption{\label{IVn} \textbf{Current-driven breakdown of crystalline order near the superconducting phase.} 
Left panel: $dV/dI$ plotted as a function of $I_{\mathrm{d.c.}}$ and $n$ in the vicinity of the superconducting phase. Right panels: $dV/dI$ plotted as a function of $I_{\mathrm{d.c.}}$ at fixed $n$ values indicated by vertical arrows in the left panel. These measurements reveal pronounced dips in $dV/dI$ at both positive and negative $I_{\mathrm{d.c.}}$, characteristic of current-driven breakdown of the crystalline order adjacent to the superconducting phase. This provides strong evidence that superconductivity coexists with an underlying CDW order. Notably, the current-driven breakdown of the crystal occurs at larger $I_{\mathrm{d.c.}}$ than the critical supercurrent, further supporting that the superconducting phase is bounded by the stability window of the CDW order. The slight asymmetry observed in the $I$--$V$ curves may arise from current-induced heating.
}
\end{figure*}

\begin{figure*}
\includegraphics[width=0.95\linewidth]{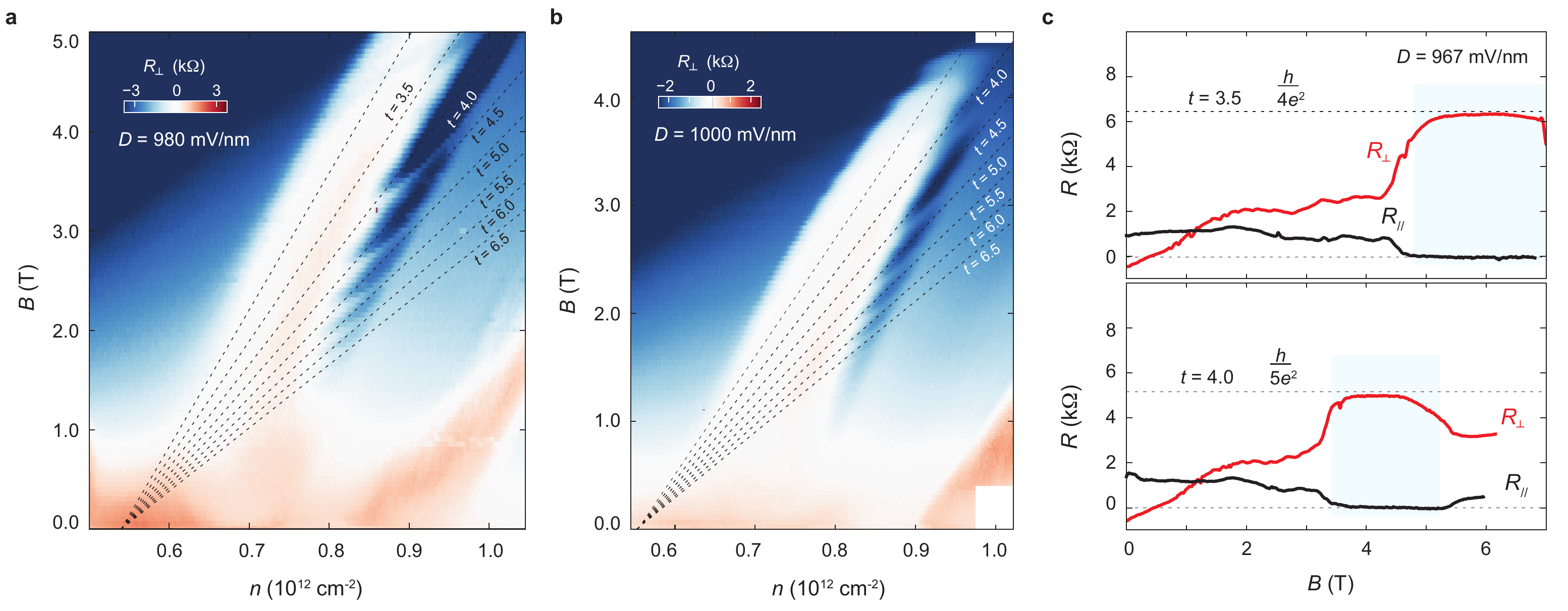}
\caption{\label{mismatch} \textbf{St\v{r}eda slopes of RIQH states and the confinement near $n/n_{\mathrm{CDW}}=2$.} 
(a,b) Maps of $R_{\perp}$ as a function of carrier density and perpendicular magnetic field measured at (a) $D=980$ mV/nm and (b) $D=1000$ mV/nm. Black dashed lines are fits to the RIQH trajectories using St\v{r}eda slopes $t=N/2$, where $N$ is an integer. Remarkably, this unconventional fit captures the entire sequence of RIQH states, in clear contrast to conventional re-entrant quantum Hall states. Despite these unconventional trajectories, the Hall response remains quantized to integer values. (c) Longitudinal resistance, \Rpara\ (black), and Hall resistance, \Rperp\ (red), measured across two representative RIQH states. The extent of the RIQH states is highlighted by the light-blue shaded regions, where vanishing \Rpara\ coincides with Hall plateaus at integer quantization.
The coexistence of half-integer St\v{r}eda slopes and integer Hall quantization reveals a pronounced mismatch between the apparent St\v{r}eda slope and the re-entrant Hall plateau. This mismatch arises naturally from the coexistence of localized and itinerant electrons. As shown in Fig.~\ref{fig2}e, $n_{\mathrm{CDW}}$ decreases with increasing $n$ near the RIQH regime, implying that the density of itinerant electrons grows faster than the total carrier density. Consequently, changes in $n$ are amplified in the itinerant sector that determines the Hall response, leading to apparent St\v{r}eda slopes that differ from those expected from the quantized Hall plateau.
}
\end{figure*}

\begin{figure*}
\includegraphics[width=0.9\linewidth]{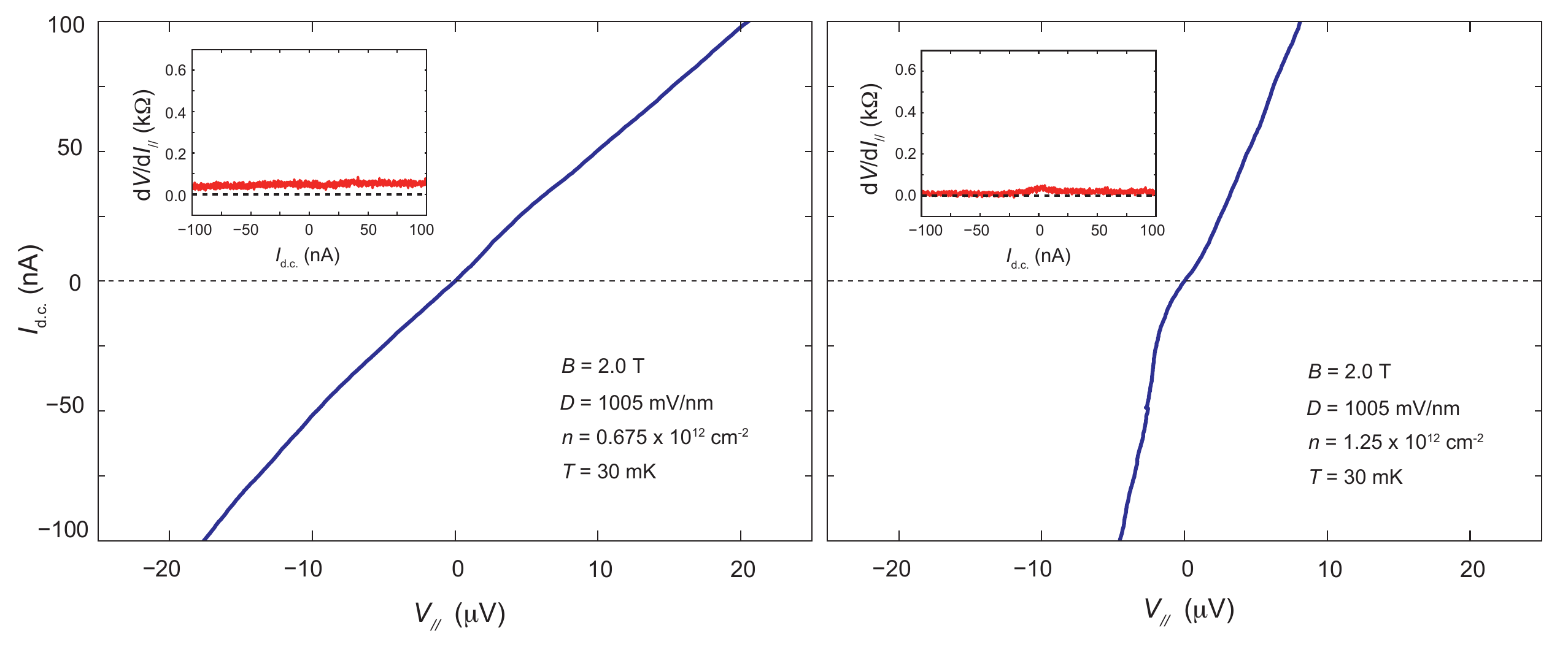}
\caption{\label{linearIV}\textbf{Linear $I$--$V$ curves outside of regimes I--III.} 
$I$--$V$ characteristics measured at $D = 1005$ mV/nm and $n= 0.675\times10^{12}$ cm$^{-2}$ (left), and $n= 1.25\times10^{12}$ cm$^{-2}$ (right), which are located outside of regimes I--III. The measurement is performed at $B = 2$ T and $T = 30$ mK. 
}
\end{figure*}

\begin{figure*}
\includegraphics[width=0.82\linewidth]{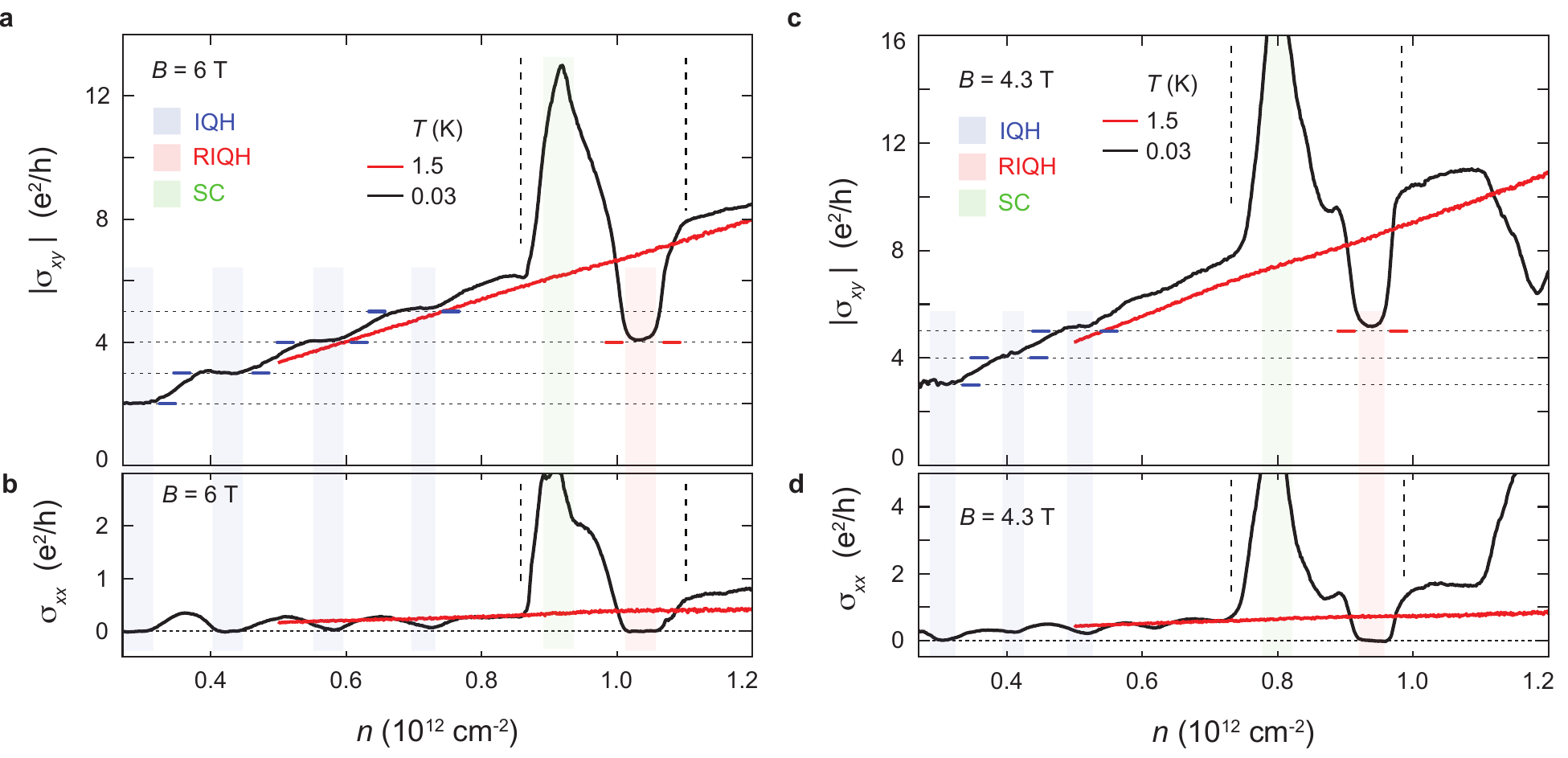}
\caption{\label{R_n_temp} \textbf{Transport response above the melting transition of the charge crystal.} 
(a,c) Hall conductivity $\sigma_{xy}$ and (b,d) longitudinal conductivity $\sigma_{xx}$ plotted as a function of $n$, measured at (a,b) $B_{\perp} = 6$~T and (c,d) $B_{\perp} = 4.3$~T. Black and red traces correspond to measurements at $T = 30$~mK and $T = 1.5$~K, respectively. The temperature $T = 1.5$~K lies above the melting transition of the CDW orders. At this temperature, signatures of regime II vanish, and the transport response within this region becomes indistinguishable from that of the surrounding metallic phase. In this regime, $\sigma_{xy}$ varies monotonically with $n$, while $\sigma_{xx}$ remains weakly dependent on $n$, consistent with conventional metallic transport. This observation supports the interpretation that regime II originates from the formation of CDW orders below $T \sim 0.5$~K.
}
\end{figure*}

\begin{figure*}
\includegraphics[width=0.62\linewidth]{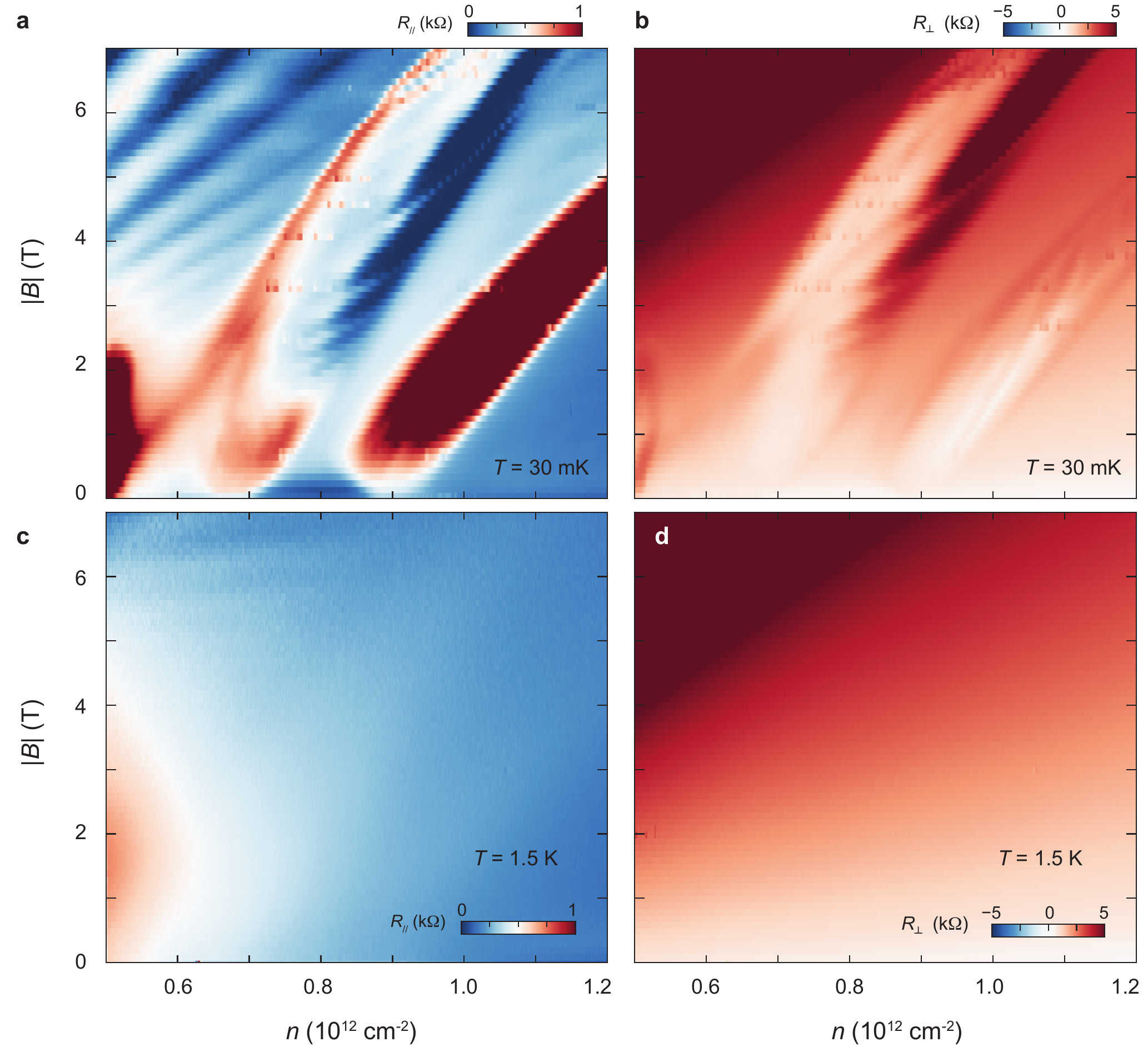}
\caption{\label{PnBT} 
\textbf{Evolution of the $n$--$B$ maps above the melting transition at 1.5~K.} 
(a,b) $n$--$B$ maps of (a) $R_{\parallel}$ and (b) $R_{\perp}$ measured at $T = 30$~mK, showing a series of RIQH states within regime II. 
(c,d) $n$--$B$ maps of (c) $R_{\parallel}$ and (d) $R_{\perp}$ measured at $T = 1.5$~K, above the melting transition of the charge crystal. At this temperature, signatures of regime II, together with the associated RIQH states, vanish.
}
\end{figure*}

\begin{figure*}
\includegraphics[width=0.65\linewidth]{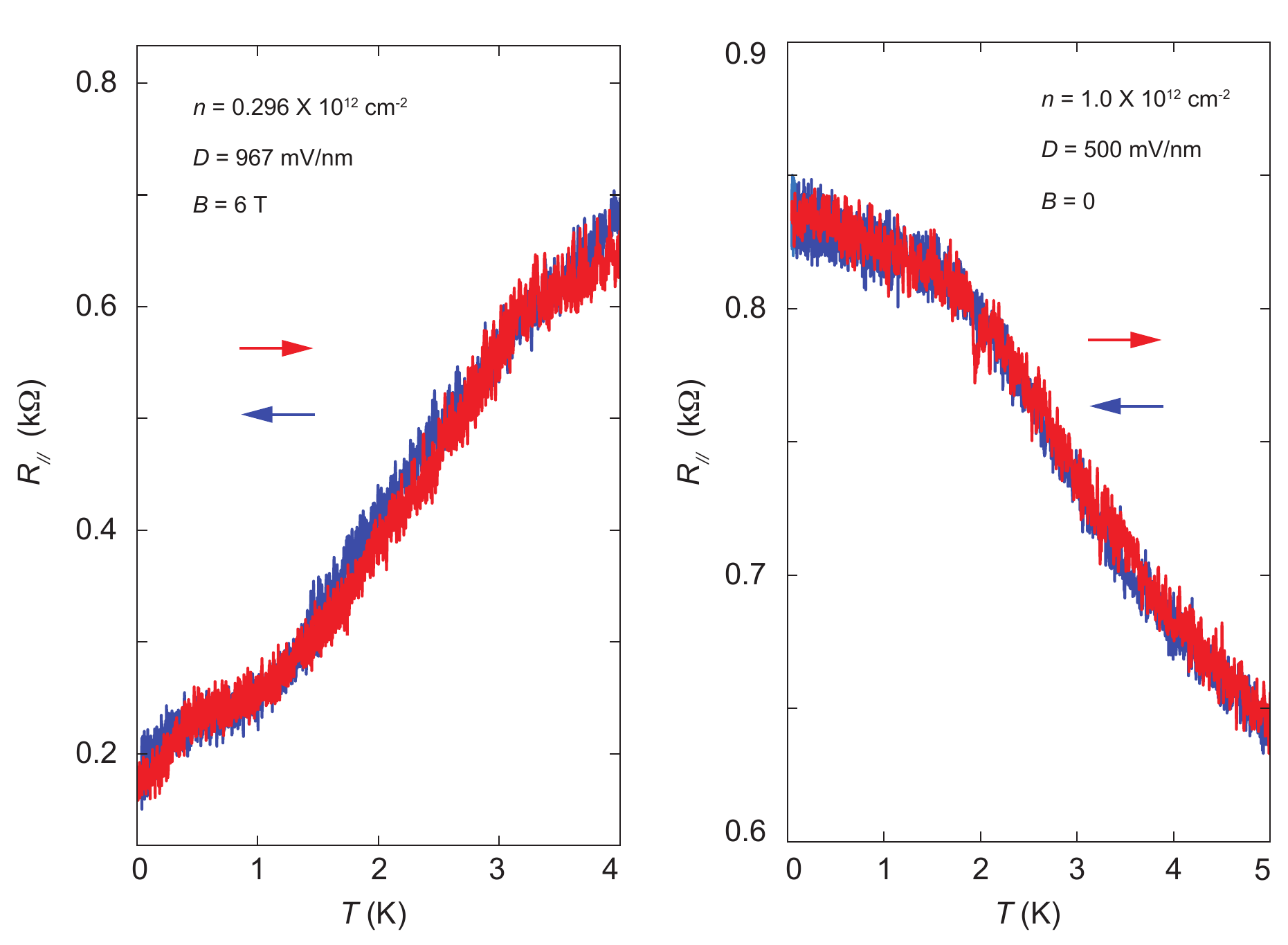}
\caption{\label{No_hysteresis}\textbf{Absence of thermal hysteresis outside the CDW regimes.} 
While pronounced thermal hysteresis is observed across regimes I--III, it is absent outside these regimes. Here, we plot \Rpara\ versus $T$ measured during cooling and warming. 
The left panel shows data taken at an $(n,D)$ value near regime II, but within a conventional IQH state at $6$ T. The right panel shows data measured at zero field in an unpolarized metallic state at a much lower $D$. In both cases, thermal hysteresis is clearly absent, providing a sharp contrast to the $R$--$T$ traces measured within regimes I--III, which exhibit pronounced thermal hysteresis indicative of first-order melting.
}
\end{figure*}

\begin{figure*}
\includegraphics[width=0.9\linewidth]{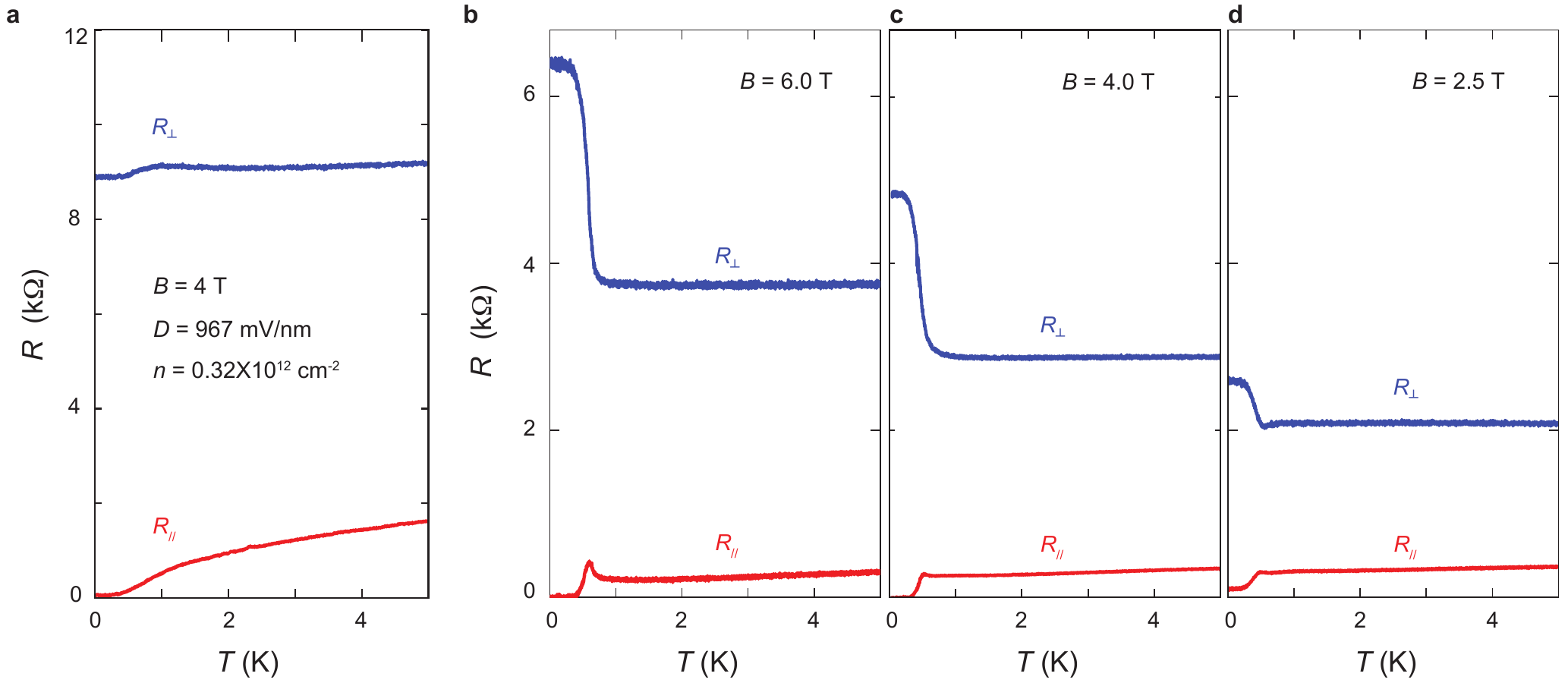}
\caption{\label{RT_QHE} 
\textbf{Temperature dependence of conventional IQH and RIQH states.} 
Temperature dependence of \Rpara\ (red traces) and \Rperp\ (blue traces) measured in 
(a) a conventional IQH state emerging from charge neutrality, and 
(b) RIQH states inside regime~II at different values of $B_{\perp}$. 
The conventional IQH state exhibits a gradual temperature evolution: \Rpara\ decreases smoothly toward zero upon cooling, while \Rperp\ remains largely temperature independent. This behavior is expected because the carrier density participating in the conventional quantum Hall state does not change with temperature. 
By contrast, both transport channels exhibit abrupt temperature-driven transitions in the RIQH states. In particular, \Rperp\ drops sharply upon warming, consistent with melting of the underlying charge crystal, which releases localized carriers, increases the itinerant carrier density, and thereby suppresses the Hall resistance.
}
\end{figure*}

\begin{figure*}
\includegraphics[width=0.95\linewidth]{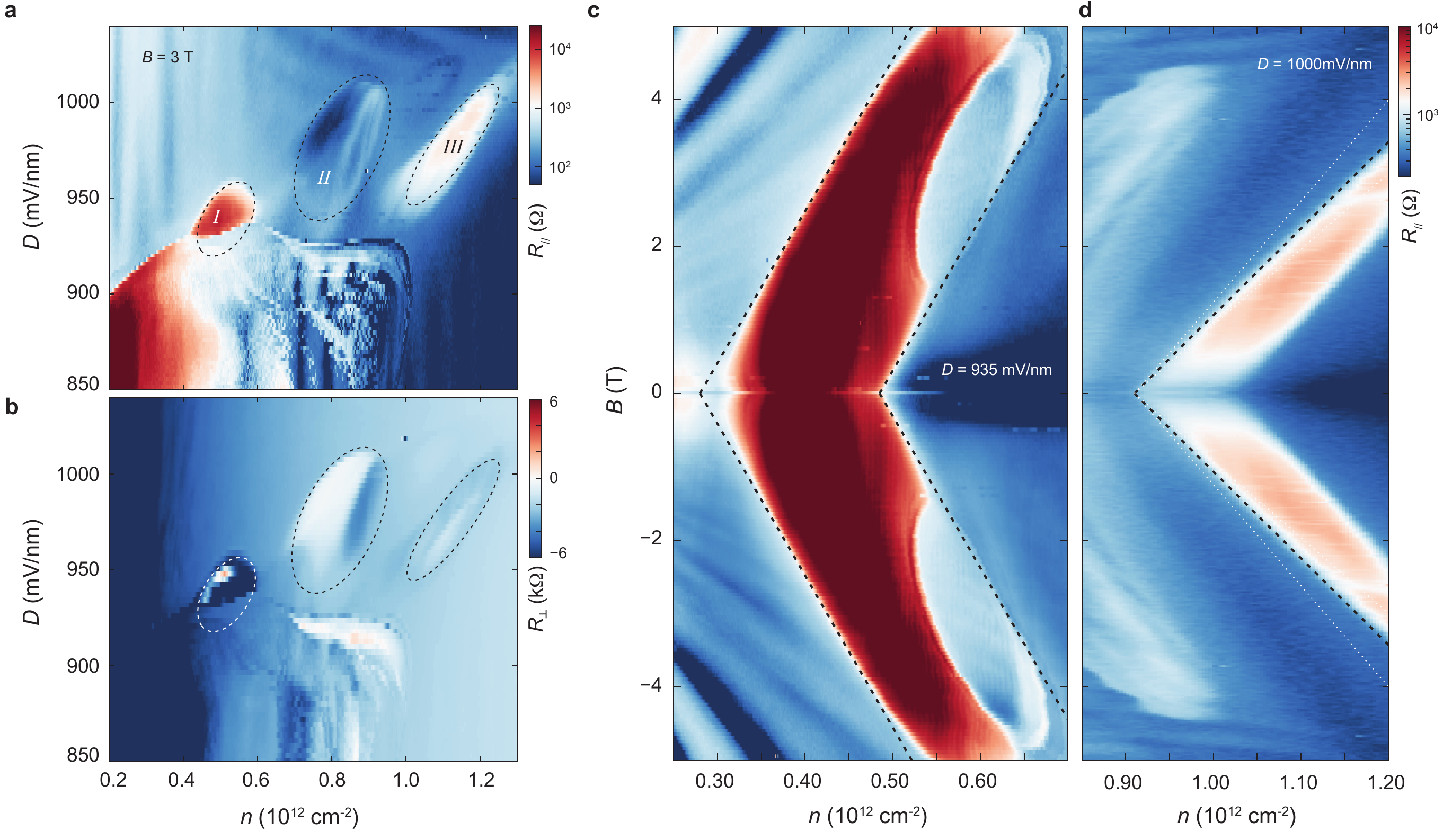}
\caption{\label{pocket}
\textbf{$n$--\Bperp\ maps of regimes I and III.}
(a,b) $n$--$D$ maps of (a) \Rpara\ and (b) \Rperp\ measured at \Bperp$=3$ T, showing regimes I--III outlined by black dashed lines.
(c,d) \Rpara\ as a function of carrier density $n$ and perpendicular magnetic field \Bperp\ measured around (c) regime I and (d) regime III. Each regime is distinguished from the surrounding metallic state by its enhanced resistance, shown as red in the chosen color scale. Remarkably, the boundaries of these regimes shift systematically with increasing \Bperp, following the trajectories indicated by the black dashed lines.
}
\end{figure*}

\begin{figure*}
\includegraphics[width=0.9\linewidth]{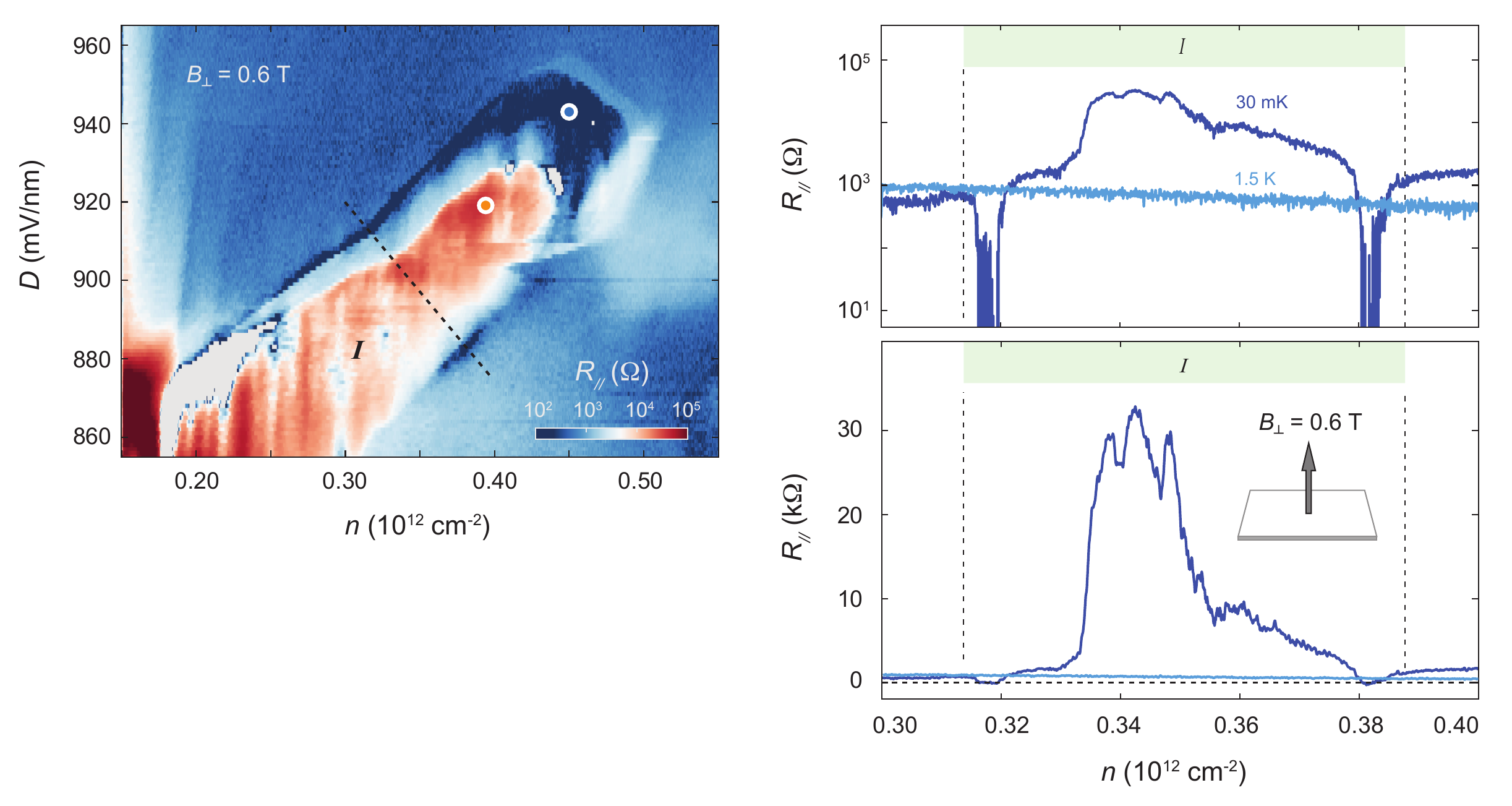}
\caption{\label{linecut} \textbf{Superconductivity in regime I.} 
Line cut along the black dashed line in the left panel is shown as line traces in the right panels. Top and bottom panels on the right show the same line traces with log scale (top) and linear scale (bottom) in the y axis. Dark and light blue traces correspond to measurements at $T=30\,\mathrm{mK}$ and $1.5\,\mathrm{K}$, respectively.
}
\end{figure*}

\begin{figure*}
\includegraphics[width=0.73\linewidth]{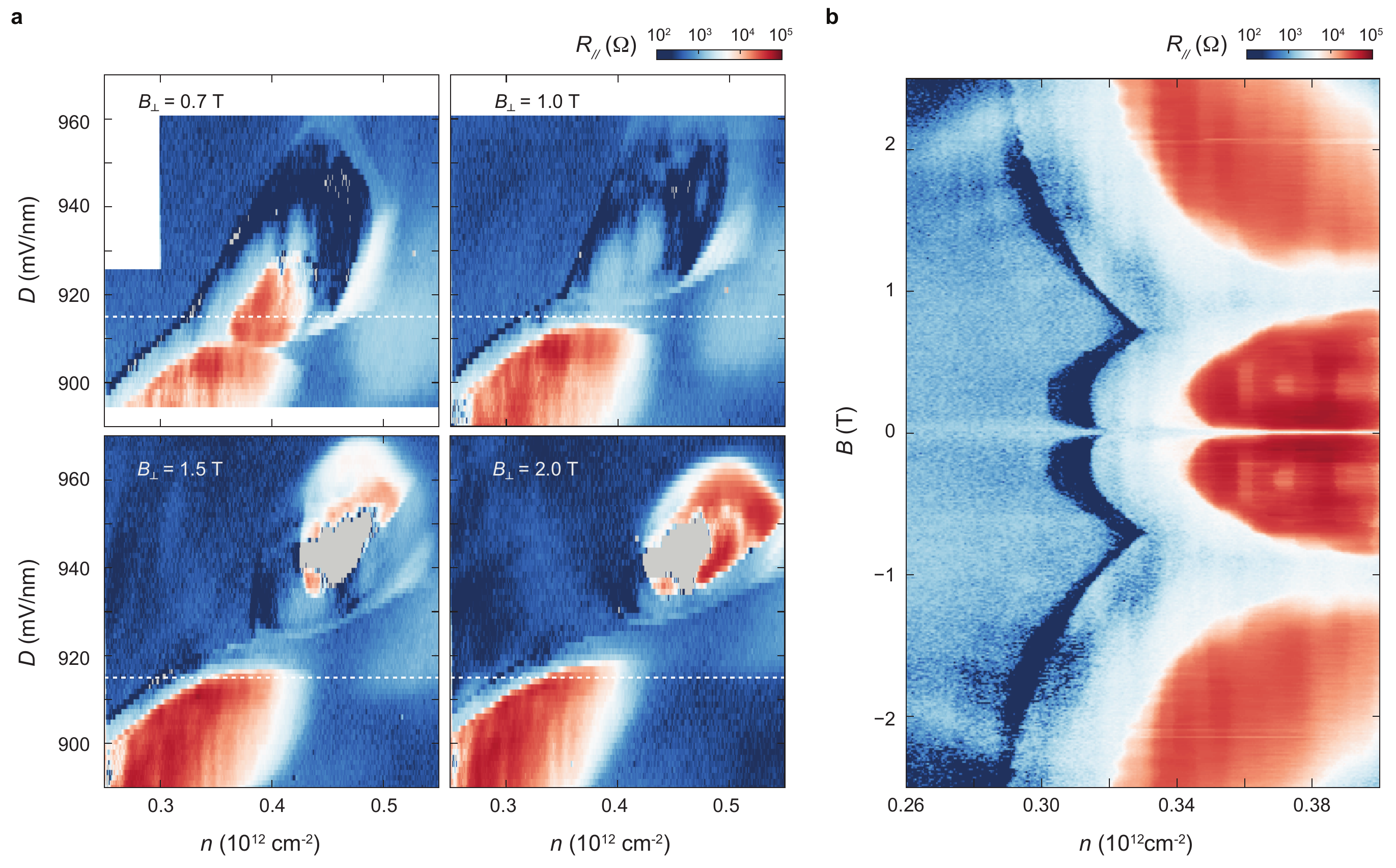}
\caption{\label{nB_SF2} 
\textbf{$n$--$B$ evolution of regime~I.} 
(a) \Rpara\ as a function of $n$ and $D$ measured around regime~I at different values of \Bperp. With increasing \Bperp, a resistive island progressively separates from the main resistive region.  
(b) \Rpara\ as a function of $n$ and $B$ measured along the white dashed line in panel (a). This $n$--$B$ map reveals a superconducting phase, evidenced by vanishing \Rpara, located on the low-density side of the resistive state. The resistive region is suppressed around $B_{\perp}\sim1$~T, corresponding to the formation of the isolated resistive island. The superconducting phase closely follows the low-density boundary of the resistive state, indicating an intimate connection between these two phases.
}
\end{figure*}

\begin{figure*}
\includegraphics[width=0.5\linewidth]{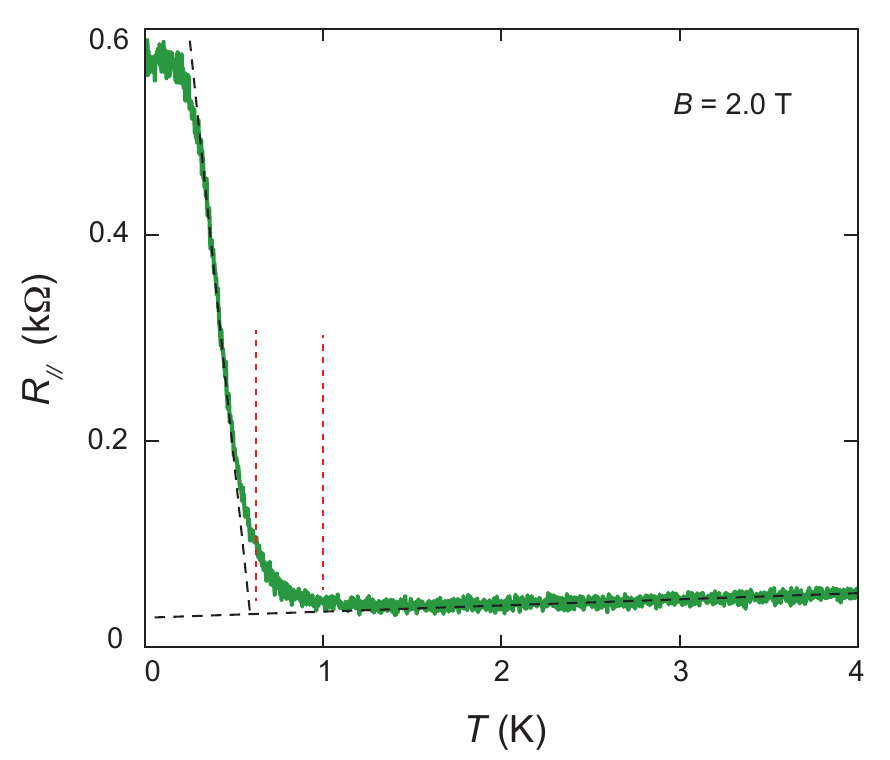}
\caption{\label{Insulating_onset}\textbf{Definition of the onset of insulating behavior.} 
$R$--$T$ trace measured from the resistive state inside regime II. The onset temperature is determined using two criteria: 
(i) the temperature at which \Rpara\ first deviates from the high-temperature trend, marked by the vertical red dashed line at higher temperature; 
(ii) the temperature obtained by extrapolating the low-temperature insulating response back to the high-temperature behavior, defined by the intersection of the two black dashed lines and marked by the vertical red dashed line at lower temperature. 
These two criteria define the uncertainty range used to determine the onset temperature shown in Fig.~\ref{onset}a.
}
\end{figure*}

\begin{figure*}
\includegraphics[width=0.85\linewidth]{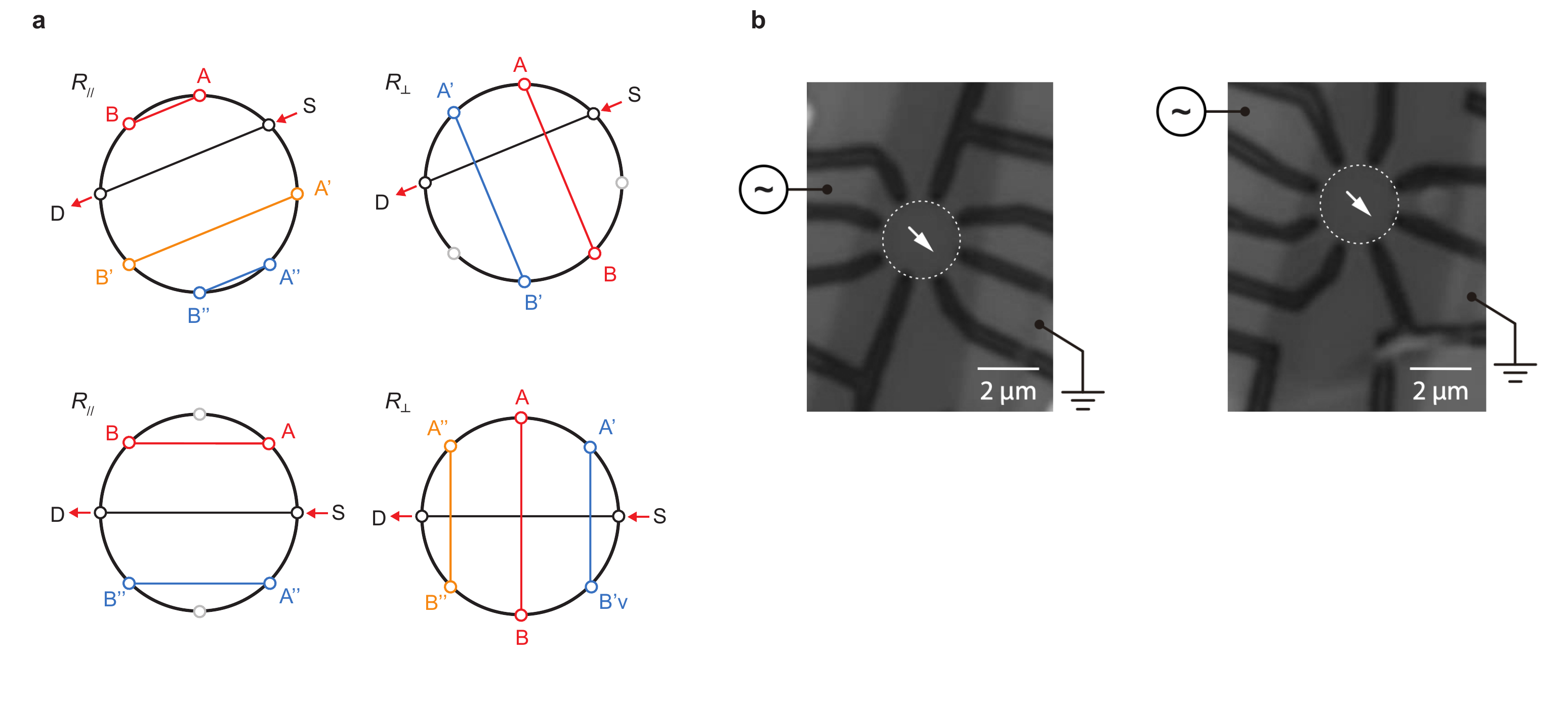}
\caption{\label{sample} \textbf{Measurement configurations.} 
(a) Schematic diagram of measurement configurations. In the left (right) panels, voltage is measured between contacts aligned parallel (perpendicular) to the current flow direction. In this manuscript, these measurements are labeled \Rpara\ and \Rperp, respectively.
In the quantum Hall regime, \Rpara\ and \Rperp\ are consistent with the longitudinal and transverse resistances, $R_{xx}$ and $R_{xy}$. This is perfectly illustrated by the results shown in Fig.~\ref{fig1}e--f and Fig.~\ref{fig2}a--d.  
(b) Optical image of two R6G samples used in this work. 
}
\end{figure*}

\begin{figure*}
\includegraphics[width=0.85\linewidth]{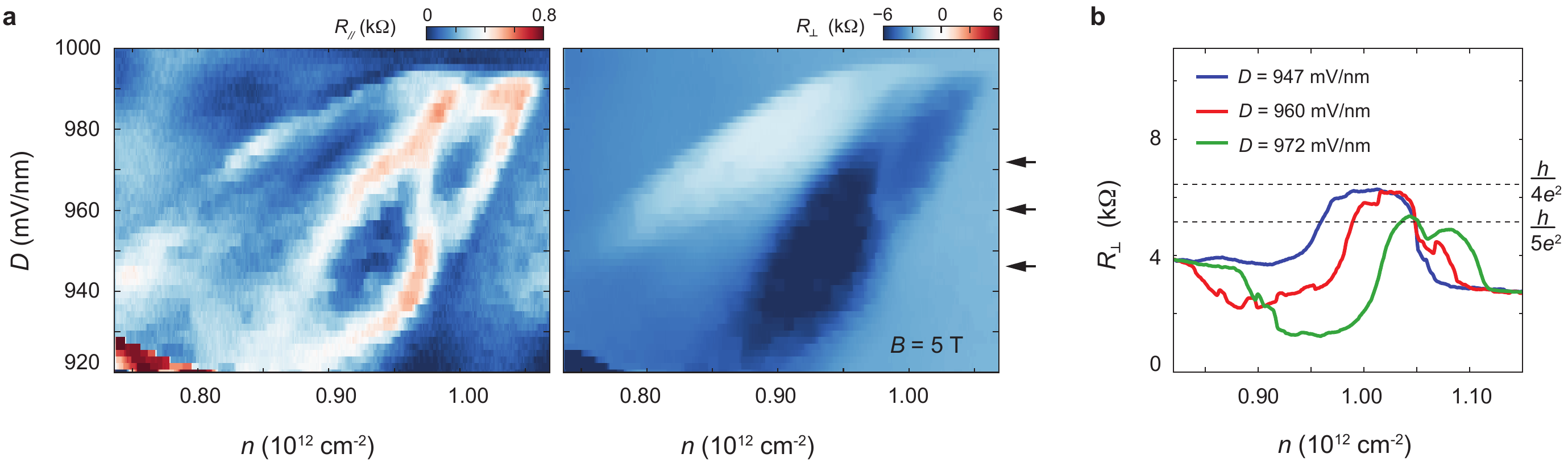}
\caption{\label{5T}{\bf{RIQH states with different Hall plateaus.}} 
(a) \Rpara\ (left) and \Rperp\ (right) as functions of $n$ and $D$ across regime~II, measured at $B_{\perp} = 5$~T. This $n$--$D$ map captures two RIQH states simultaneously. The boundary of each state is defined by a resistance peak in \Rpara, while the two states exhibit distinct quantized Hall plateaus.  
(b) \Rperp\ as a function of $n$, taken at constant values of $D$ from (a). Blue and red traces, measured at $D = 947$ and $960$~mV/nm, respectively, show Hall plateaus near $h / 4e^{2}$. The green trace, taken at $D = 972$~mV/nm, exhibits a plateau near $h / 5e^{2}$.  
}
\end{figure*}

\begin{figure*}[h]
\includegraphics[width=0.9\linewidth]{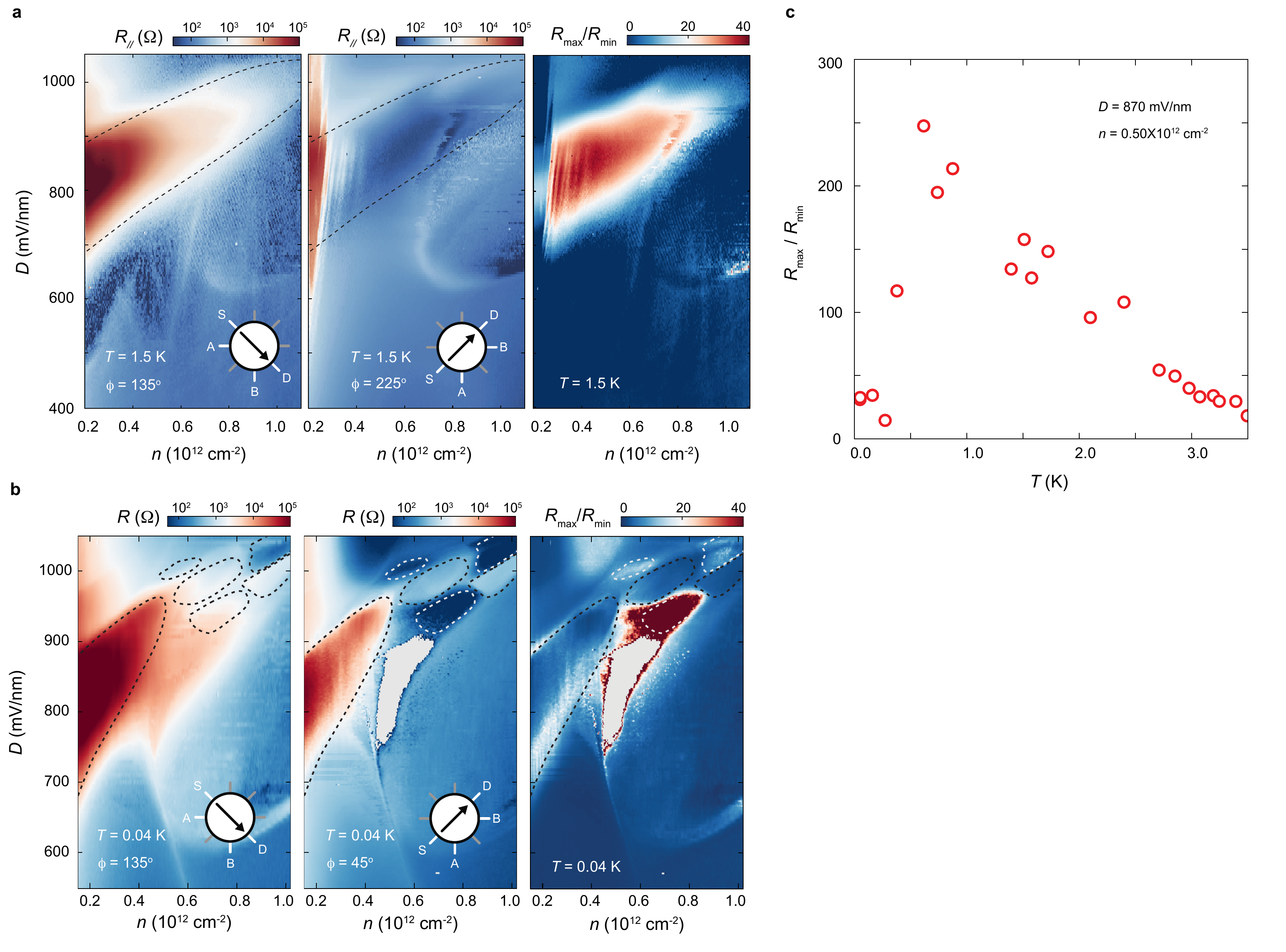}
\caption{\label{nD_anisotropy} \textbf{Transport anisotropy across the $n$--$D$ phase space.} 
\Rpara\ measured with current flowing along $\phi = 135^{\circ}$ (left panel) and $\phi = 45^{\circ}$ (middle panel), and the ratio $R_{\mathrm{max}}/R_{\mathrm{min}}$ (right panel), plotted as a function of $n$ and $D$ at \Bperp\ $=0$. Measurements are performed at (a) $T = 1.5$~K and (b) $T = 40$~mK. At $T = 1.5$~K, the entire branch of the $n$--$D$ phase space exhibits uniform anisotropy, as evidenced by the large values of $R_{\mathrm{max}}/R_{\mathrm{min}}$, consistent with the emergence of a stripe or nematic state~\cite{Qin2025stripeSC}. At $T = 40$~mK, the emergence of CDW orders in regimes I–III leads to a pronounced suppression of transport anisotropy, while the response remains anisotropic in the vicinity of the SC\,i phase. (c) Temperature dependence of $R_{\mathrm{max}}/R_{\mathrm{min}}$ measured within regime I, revealing a transition near $T \sim 0.5$~K, in excellent agreement with the onset temperature of the Wigner-solid-like order. Together, these observations suggest that the CDW phase emerges from a higher-temperature anisotropic state, while exhibiting substantially reduced anisotropy in the crystalline phase.
}
\end{figure*}

\end{widetext}

\end{document}